\documentclass{aa}  
\usepackage{natbib}
\bibpunct{(}{)}{;}{a}{}{,}
\usepackage{graphicx}
\usepackage{txfonts}
\usepackage{float}
\usepackage{amsmath}	
\usepackage{booktabs}
\usepackage{comment}
\usepackage{subcaption}
\usepackage{pdflscape}
\usepackage{newtxtext,newtxmath}
\usepackage[T1]{fontenc}
\usepackage{hyperref}
\hypersetup{
  colorlinks   = true,   
  urlcolor     = blue,    
  citecolor    = blue      
}

\newcommand{\hi}{\textsc{H\,i}\space}
\newcommand{\kms}{$\,$km$\,$s$^{-1}$}

\begin{document}

   \title{A MeerKAT study of a neutral hydrogen rich grouping of galaxies with megaparsec-scale filamentary-like structure}

   \author{G. D. Lawrie \inst{1}
          \and
          R. P. Deane\inst{1,2}
          \and 
          R. Davé \inst{3,4,5}
          }

   \institute{Wits Centre for Astrophysics, School of Physics, University of the Witwatersrand,
              1 Jan Smuts Avenue, Johannesburg, 2000, South Africa\\
         \and
             Department of Physics, University of Pretoria, Private Bag X20, Pretoria 0028, South Africa\\
        \and
            Institute for Astronomy, University of Edinburgh, Royal Observatory, Blackford Hill, Edinburgh EH9 3HJ, UK.\\
        \and
            University of the Western Cape, Department of Physics and Astronomy, Bellville, Cape Town 7535, South Africa\\
        \and
            South African Astronomical Observatory, Observatory, Cape Town 7925, South Africa}

   \date{Received May 30, 2025; accepted July 18, 2025}
 
  \abstract 
   {Environmental effects within cosmological overdensities, such as galaxy groups and clusters, have been shown to impact galaxies and their cold gas reservoirs and thereby provide constraints on galaxy evolution models. Galaxy groups foster frequent galaxy-galaxy interactions, making them rich environments in which to study galaxy transformation.}
   {In this work, we study a serendipitously discovered large overdensity of \hi galaxies at $z\sim0.04$. The galaxies appear to lie in a filamentary-like structure of megaparsec scale. MeerKAT's angular resolution and field of view allow us to spatially resolve the \hi galaxies while simultaneously probing large-scale structure.}
   {The \hi and sub-arcsec Dark Energy Survey (DES) imaging reveal a large number of interacting/disturbed galaxies in this collective group. MeerKAT data enables us to derive \hi masses and investigate interacting galaxies. We use DES and Wide-field Infrared Survey Explorer (WISE) data to quantify the star formation rates, stellar masses, and stellar morphologies of member galaxies and compare these with field scaling relations. To place this discovery and the environmental effects in context, we use the \textsc{Simba} cosmological hydrodynamical simulation to investigate the prevalence of qualitatively similar \hi overdensities and their large-scale morphological properties. This enables a prediction of how frequently such structures might be serendipitously discovered with MeerKAT and SKA-Mid \hi observations in comparable observation time.}
   {The combination of spatially resolved \hi data and optical imaging reveals a group rich in interactions, suggesting environmental processes are already shaping galaxy properties within the structure.}
   {More of these serendipitous discoveries are expected and, alongside ongoing targeted programmes, these will provide a rich, unbiased sample to study galaxy transformation and enable a MeerKAT \hi perspective on large-scale structure, including filaments.}

   \keywords{Galaxies: evolution --
                Galaxies: formation --
                Galaxies: interactions --
                Galaxies: groups: general --
                Radio lines: galaxies --
                (Cosmology:) large-scale structure of Universe
               }

    \titlerunning{Filamentary galaxy group}
    \authorrunning{G. D. Lawrie et al.}
   \maketitle

\section{Introduction}

Currently, the most widely accepted cosmological structure formation paradigm is the $\Lambda$ cold dark matter ($\Lambda$CDM) model, which predicts that all galaxies form and evolve within dark matter halos and dark energy accelerates expansion at late cosmological epochs \citep{Springel_2005}. Galaxy groups have been shown to form an integral part of this hierarchical evolution model \citep{Springel_2018}, making the study of these groups important for our understanding of large-scale structure formation.

In galaxy clusters, environmental effects such as tidal interactions \citep{Moore_1998}, and ram-pressure stripping \citep{Gunn_1972} are commonly observed. This is due to the high galaxy number density leading to an increase in galaxy-galaxy interactions and the hot and dense intracluster medium (ICM) providing a medium for gas stripping. Galaxy groups are generally less dense and have fewer members \citep[$\sim3-100$,][]{Catinella_2013} than their larger cluster counterparts ($>100$ members), with lower dark matter halo masses of $\sim10^{12-14}\,\rm{M}_{\odot}$ \citep{Catinella_2013,Hess_2013}. However, it is important to note the difference between clusters and high-mass groups is not well defined and no clear distinction exists in the literature \citep[see][]{Lovisari_2021}. Albeit less common, tidal interactions are also observed in lower halo mass group structures, both in compact groups of $\sim4$ galaxies with $\sim30-40\,\rm{kpc}$ projected separation \citep[eg.][]{Sulentic_2001} as well as in larger groups of $\sim20$ members \citep[eg.][]{Ranchod_2021}. There is evidence of these interactions in group-like environments before in-fall into a cluster environment (often referred to as "pre-processing"), such as in the Fornax A group \citep{Kleiner_2021,Loubser_2024}. About half of all optically identified galaxies reside in group environments at $z=0$ \citep{Eke_2004,Robotham_2011}, and 25$\,\%$ of neutral hydrogen (\textsc{H\,i}) identified galaxies reside in group environments \citep{Hess_2013}, making the study of galaxies in this environment a key path to improving our understanding of galaxy evolution.

It has been shown that the interstellar medium (ISM) properties of galaxies are intimately linked to the circumgalactic medium \citep{Tumlinson_2017,Nielsen_2020}. The \hi in galaxies is typically significantly more extended than the stellar component \citep{Broeils_1997,Leroy_2008}. This, together with the sensitivity of \hi to tidal interactions and ram-pressure stripping, makes \hi observations a powerful tool for probing the influence of environmental effects as a tracer of long dynamical timescales, as can be seen in the M81, M82 and NGC3077 triplet \citep{Yun_1994}. \hi has also been used as a tool to probe cold gas replenishment from large-scale structure such as filaments \citep{Kleiner_2017}. The \hi mass function \citep[HIMF;][]{Jones_2018} and the $M_{\rm{HI}}-M_{*}$ relation \citep{Parkash_2018} provide important probes of environmental effects. However we require a large number (of order hundreds) of galaxies to ensure sufficient statistical significance in a given environment.

MeerKAT \citep{Jonas_2016}, an SKA-mid precursor, has already made significant contributions to the exploration of the \hi universe through surveys such as: Looking At the Distant Universe with the MeerKAT Array \citep[LADUMA;][]{Blyth_2016}, the \hi component of the MeerKAT International GigaHertz Tiered Extragalactic Exploration survey \citep[MIGHTEE;][]{Jarvis_2016,Maddox_2021,Ranchod_2021,Heywood_2024}, MeerKAT \hi Observations of Nearby Galactic Objects: Observing Southern Emitters \citep[MHONGOOSE;][]{Blok_2024}, and the MeerKAT Absorption Line Survey \citep[MALS;][]{Gupta_2016,Deka_2024}.

The advances in both cosmological hydrodynamical simulations and \hi observations over the past years provide an opportunity for comparison between empirical and theoretical results. \textsc{Simba} \citep[][]{Dave_2019}, \textsc{EAGLE} \citep[][]{Schaye_2015}, and \textsc{IllustrisTNG} \citep[][]{Pillepich_2018} have been shown to reproduce the HIMF \citep[see ][for a comparison]{Dave_2020}. However, the environmental effects on \hi within large-scale structures have not been studied as well \citep[e.g.][]{Bulichi_2024}, so it is valuable to explore additional constraints toward this.

In this paper, we perform a detailed study of the serendipitous discovery of a large grouping of \textsc{H\,i}-rich galaxies in the foreground of the Abell~3365 galaxy cluster, which we first reported in \citet{Knowles_2022}. The grouping contains 20 detections and takes on a peculiar filamentary-like shape in projection. The galaxies span a projected $\sim 3.7 \, \rm{Mpc}$ from the northernmost to the southernmost galaxy and include several interacting systems. In \citet{Knowles_2022}, we showed a sample of two \hi galaxies with radio continuum counterparts and an overview \hi map of the group before group member identification. Here we perform individual galaxy analysis of the \hi galaxies detected in the grouping using multiwavelength data from the Dark Energy Survey (DES), an optical imaging survey carried out by the Dark Energy Camera \citep[DECam, ][]{Abbott_2021}, and \textsc{NASA's} Wide-field Infrared Survey Explorer (WISE) \citep{Wright_2010}, together with MeerKAT data to probe \hi and stellar masses, kinematics, galaxy-galaxy interactions, and optical morphologies. We use both manual and automated source finding, along with a Friends-of-Friends algorithm, to establish robust group membership. We then explore similar structures in \textsc{Simba} to contextualise this \hi overdensity. 

This paper is organised as follows: In Sect.~\ref{sec:cali} we describe the MeerKAT observations, the calibration and imaging implemented in this work, and summarise the ancillary data. In Sect.~\ref{sec:results}, the source-finding process is discussed, and we then briefly address the approach we take to group member selection. We also present moment 0, moment 1, DES \textit{rgb}, WISE, and radio continuum cutouts of individual galaxies. In Sect.~\ref{sec:discussion} we discuss the morphology and kinematics of individual galaxies as well as potentially interacting systems within the grouping. The properties of the group are then compared to established scaling relations. Finally, we use the \textsc{Simba} cosmological hydrodynamical simulations to put the peculiar filamentary-like projection of the grouping into context with respect to the large-scale structure. A summary of the findings of this work is then presented in Sect.~\ref{sec:conclusion}. Throughout this paper we adopt cosmological values found in \citet{Planck_2020}, $\Omega_M = 0.315 \pm 0.007$ and $H_0 = 67.4 \pm 0.5 \,\rm{km}\,\rm{s}^{-1}\,\rm{Mpc}^{-1}$. This corresponds to a scale of $0.822\,\rm{kpc}\,\rm{arcsec}^{-1}$ for sources at $z=0.04$.

\section{Data, calibration, and ancillary data}
\label{sec:cali}

\subsection{MeerKAT observations}
\label{mgcls}

MeerKAT is a radio interferometer located in the Northern Cape of South Africa. The mid-frequency array serves as a precursor to the Square Kilometre Array (SKA). The MeerKAT array consists of 64 antennas in a dense core layout where $\sim70\,\%$ of the antennas are in the core with a minimum baseline of 29~m and a maximum baseline of $\sim1\,\rm{km}$. Approximately $30\,\%$ of the antennas are located beyond the core and provide the interferometer with a maximum baseline of $\sim8$~km. Both regions are distributed such that they provide near Gaussian $(u,v)$ coverage. All dishes are 13.5~m in diameter with offset Gregorian optics and equipped with S-band, L-band, and UHF-band receivers with frequency ranges of $1750 - 3500 \, \mathrm{MHz}$, $900 - 1670 \, \mathrm{MHz}$, and $580 - 1015 \,  \mathrm{MHz}$ respectively. For detailed technical specifications, see \citet{Jonas_2009, Jonas_2016, Camilo_2018}.

The raw visibility data used in this paper stem from the MeerKAT Galaxy Cluster Legacy Survey (MGCLS), a SARAO Science Legacy Survey that performed deep MeerKAT L-band observations of 115 galaxy clusters \citep{Knowles_2022}. Each cluster was typically observed for 6–10 hours each with MeerKAT in the correlator's 4k configuration, which has 4096 channels leading to a channel width of $209\,\rm{kHz}$ (corresponding to $\sim 44$\kms at $z = 0$).

The group lies in the Abell~3365 field, located at $\rm{RA}=05\rm{h}48\rm{m}12\rm{s}$ and $\rm{Dec}=-21\rm{d}56{m}03{s}$. Note that Abell~3365 is at $z\sim0.093$, within a frequency range affected by radio frequency interference (RFI), and is cosmologically unrelated to the galaxies studied in this work. Both observations used in this work had a correlator dump time of 8 seconds and included 3 sources: J0408-6545 (bandpass calibrator at $\rm{RA}=04\rm{h}08\rm{m}20\rm{s}, \rm{Dec}=-65\rm{d}45\rm{m}07\rm{s}$), J0609-1542 (complex gain calibrator at $\rm{RA}=06\rm{h}09\rm{m}40\rm{s}$, $\rm{Dec}=-15\rm{d}42\rm{m}40\rm{s}$), and Abell~3365. The observations' properties are summarised in Table~\ref{tab:obs_props}.

\begin{table}
    \centering
    \caption{Summary of MeerKAT observations.}
    \label{tab:obs_props}
    \renewcommand{\arraystretch}{1.2}
    \begin{tabular}{llll}
        \hline
        Experiment ID & Date & Antennas & Track Length \\
        \hline
        20181114-0020 & 2018/11/14-15 & 56/64 & 10.05 hours \\
        20181118-0008 & 2018/11/18-19 & 59/64 & 10.10 hours \\
        \hline
    \end{tabular}
\end{table}

\subsection{Calibration and imaging}
\label{sec:calibration}

The MeerKAT raw visibilities are reduced using \textsc{Oxkat}\footnote{https://github.com/IanHeywood/oxkat} \citep{Heywood_2020} through a \textsc{Singularity} container on \textsc{ILIFU} high-performance computing cluster, which is developed and maintained by \textsc{IDIA}\footnote{https://www.idia.ac.za/}. \textsc{Oxkat} is a semi-automatic complete data reduction pipeline software that implements standard routines such as flagging, reference calibration (1GC), self-calibration (2GC), and direction-dependant calibration (3GC). \textsc{Oxkat} makes use of \textsc{Casa}\footnote{https://casa.nrao.edu} \citep{McMullin_2007}, \textsc{WSClean} \citep{Offringa_2014}, and \textsc{CubiCal}\footnote{https://github.com/ratt-ru/CubiCal} \citep{Kenyon_2018} among other software.

The autoflagger \textsc{Tricolour}\footnote{https://github.com/ratt-ru/tricolour} is used in conjunction with manual flagging of bad antennas and $(u,v)$ ranges to remove RFI and bad data. Cross calibration is done using \textsc{Oxkat} through standard \textsc{Casa} functions such as \texttt{gaincal} and \texttt{applycal} to correct for atmospheric phase shifts among other effects. J0609-1542 is observed for $\sim1$ minute every $\sim10$ minutes and used as the complex gains calibrator. We use J0408-6545 as the flux and bandpass calibrator. Self-calibration is performed to improve the fidelity and dynamic range of the image by iteratively improving the mask used by \texttt{WSClean}.

Continuum subtraction was performed in 3 steps, first in the visibility plane using \textsc{Casa} by subtracting the current sky model contained in the model data column from the corrected data using \texttt{uvsub}. Thereafter, remaining continuum emission is modelled and subtracted through fitting a 1st-order polynomial to the emission in the visibility plane at the native 8-sec resolution with \texttt{uvcontsub}. Using \texttt{WSClean} and a \texttt{Briggs} weighting scheme with a \texttt{robust} parameter of 2.0 we image the data cube. This is to produce the most sensitive data product, thereby complementing the source-finding-orientated science goals of this work. Finally, we use \texttt{imcontsub} to model and remove continuum emission in the image plane by fitting a 2nd-order polynomial to the cube channels. Polynomial fitting orders as high as 4th-order were investigated but did not lead to significant differences in the integrated spectra, moment maps or derived \hi masses, thereby having no impact on the scientific outcomes of this paper. The final sub-cube consists of 102 channels covering a frequency range of $1.3523 - 1.3734\,\rm{GHz}$, and has a restoring beam dimension of $30.6547"\times30.6547"$.

\subsection{MeerKAT data properties}

The radio continuum image used in this study is from the MGCLS raw visibilities. The continuum image is a $10240~\times~10240$ pixel image with a restoring beam of $6.87" \times 6.87"$ with a root mean square ($\rm{RMS}$) of $\sim~5\,\mu\rm{Jy}\,\rm{beam}^{-1}$. The map is centered on $\rm{RA}\ 05\rm{h}\,48\rm{m}\,12\rm{s}$ and $\rm{DEC}\ -21\rm{d}\,56\rm{m}\,06\rm{s}$. The full continuum image is relatively free of bright, troublesome sources and obvious artefacts.

The data cube produced using the method outlined in~\ref{sec:calibration} has a circular restoring beam of $30.65" \times 30.65"$ and a pixel size of $2" \times 2"$. The cube has a channel width of $209\,\rm{kHz}$, which corresponds to a velocity width of $\sim46\rm{km}\,\rm{s}^{-1}$ at $z=0.0395$. With an $\rm{RMS}$ of $\sim52\,\mu\rm{Jy}\,\rm{beam}^{-1}$, we achieve an \hi column density sensitivity of $3.146\times10^{18}\mathrm{cm}^{-2}$ at $1\sigma$. A summary of the specifications of the reduced data is given in Table~\ref{tab:cube_properties}. 

\begin{table}
    \centering
    \caption{MeerKAT \hi data cube properties.}
    \label{tab:cube_properties}
    \renewcommand{\arraystretch}{1.2}
    \begin{tabular}{p{4.2cm} p{3.5cm}}
        \hline
        Property & Value \\
        \hline
        Restoring beam & $30.65''\times30.65''$ \\
        BPA & $0^\circ$ \\
        Pixel size & $2''\times2''$ \\
        Channel width & $209\,\mathrm{kHz}$ \\ 
        & $\sim46\,\mathrm{km}\,\mathrm{s}^{-1}$ (at $z = 0.04$) \\
        Robust weighting & $2.0$ \\
        RMS & $\sim52\,\mu \mathrm{Jy}\,\mathrm{beam}^{-1}$ \\
        \hi column density sensitivity ($1\sigma$) & $3.15\times10^{18}\,\mathrm{cm}^{-2}$ \\
        \hline
    \end{tabular}
\end{table}

\subsection{Ancillary data}
\label{sec:ancillary}

For the identification of optical counterpart galaxies, we use the DES Data Release 2 (DR2). DECam is situated in Chile at Cerro Tololo Inter-American Observatory on the 4~m Blanco telescope. The data we use comes from the coadded image data products of the DES wide-area survey that covered $\sim5000$~deg$^{2}$ of the southern Galactic cap, observed in 5 bands: \textit{g, r, i, z}, and Y at 1.11", 0.95", 0.88", 0.83", and 0.90" respectively \citep{Abbott_2021}. DES has a median coadded catalogue depth of 24.7, 24.4, 23.8, 23.1, and 21.7 AB mag (for an aperture of 1.95" diameter at SNR=10), for the different bands respectively, an internal astronomical precision of $\sim$27 mas, and a photometric accuracy of $\sim$10~mag. In Fig.~\ref{fig:page1} we show the \textit{g}-band cutouts in the third column.

The mid-infrared data used in this analysis is from WISE, which mapped the entire sky between 7 January 2010 and 6 August 2010. WISE observed in the following 4 bands: W1 (3.4$\,\mu \mathrm{m}$), W2 (4.6\,$\mu \mathrm{m}$), W3 (12\,$\mu \mathrm{m}$), and W4 (22\,$\mu \mathrm{m}$) that have an angular resolution of 6.1", 6.4", 6.5", and 12.0" respectively \citep{Wright_2010}. WISE achieves 5$\sigma$ sensitivities of 0.08, 0.11, 1.00, and 6.00 mJy in unconfused regions on the ecliptic for the bands, respectively. The W1 is used to estimate $M_\ast$ for the \hi detected galaxies \citep{Jarrett_2023}, while W3 is used to calculate the $\rm{SFR}$ of these galaxies \citep{Cluver_2017}. We show a cutout of W3 in the fourth column in Fig.~\ref{fig:page1}.

The aforementioned ancillary data greatly enrich the study of the \hi group. Stellar properties of galaxies can be determined with the WISE data, which allows comparisons to scaling relations. DES images have sub-arcsecond resolution, giving us a robust way to both confirm or reject potential detection candidates from source-finding, as well as assisting in differentiating sources in the relatively poor angular resolution WISE data to avoid confusion and mismatching during cross-matching.

\section{Results}
\label{sec:results}
\subsection{Source finding}
\label{sec:source_finding}

The source finding software implemented in this work is \textsc{SoFiA~2}\footnote{https://gitlab.com/SoFiA-Admin/SoFiA-2} \citep[\textsc{Source Finding Application},][]{Westmeier_2021}. This is a fully automated source finder specifically optimised for extragalactic \hi spectral line source finding in 3D data cubes. It is based on \textsc{SoFiA} \citep{Serra_2015} with significant improvements in speed and efficiency through inclusion of \textsc{OpenMP} to allow parallelisation. These changes make \textsc{SoFiA 2} an ideal source-finding tool for this work.

The source-finding strategy is as follows: First, the default settings are used together with the algorithm recommended by the authors (S+C or smooth and clip algorithm) to establish a baseline. Thereafter, we do multiple runs of the source finder with various values for \texttt{scfind.threshold} between 3 and 5, which is the flux threshold relative to the measured noise in each smoothing iteration and the range recommended in the manual. We find that a flux threshold of 3.4 provides us with a complete catalogue when compared to a manual inspection of the cube. In this case a manual inspection refers to looking at every channel in the cube using \textsc{CARTA}\footnote{https://cartavis.org} \citep{Comrie_2021} and placing apertures on bright features, then examining the extracted spectra to classify the feature as real, noise, or continuum subtraction error. However, this rarely results in the presence of false positives in the \textsc{SoFiA 2} generated catalogue, which are manually removed. 

The majority of linking parameters are left as default with some minor adjustments to better suit our input data cube. We utilise the \texttt{linker.minSizeXY} and \texttt{linker.minSizeZ} to filter out noise peaks. We set \texttt{linker.minSizeXY} = 10 to approximate our beam size of $\sim30~\rm{arcsec}$ to discard sources smaller than the synthesized beam. We set \texttt{linker.minSizeZ} = 2, which is significantly lower than the default of 5. This is because, at a channel width of 46 \kms, 5 channels would discard all galaxies with a full velocity width of less than $\sim250$ \kms. Note that \texttt{linker.minSizeZ}~=~1 was tested and no significant difference was noticed in the final catalogue, therefore to avoid single channel noise peaks we choose to use \texttt{linker.minSizeZ} = 2. Finally, we set \texttt{linker.radiusXY} = 2 instead of the default of 1. This is done as we observe multiple interacting systems and aim to capture these interactions as opposed to separating the sources involved.

To limit the number of false positives due to a lower \texttt{scfind.threshold}, the reliability filter is used after being fine-tuned. We make use of the diagnostic plots output by \textsc{SoFiA 2} to achieve an optimal reliability filter setup. This is done by optimising the size of the kernel used to determine the reliability of the source, \citep[see][for details]{Westmeier_2021}. The goal is to obtain a normalised distribution of Skellam parameters that closely resembles a standard Gaussian with $\sigma = 1$. The \texttt{reliability.autoKernel} does not converge given a tolerance of \texttt{reliability.tolerance} = 0.1 and \texttt{reliability.iterations} = 100. However, through trial and error, we find that \texttt{reliability.scaleKernel} = 0.34 results in a distribution of normalised Skellam parameters with a median of $\mu = -0.083$ in the Skellam plot. This is a good indication that the kernel size is optimised for the data cube. Using the reliability plot, we set \texttt{reliability.minSNR} = 2.8. This is slightly below the default of \texttt{reliability.minSNR} = 3.0 to allow for potential detections closer to the noise level in the mean and summed flux density parameter space. The reliability threshold was established with \texttt{reliability.threshold} = 0.70 \citep[see][for details]{Westmeier_2021}. We note that this value is significantly lower than the default of 0.9; however, in the final catalogue, only ID 17 has a reliability of < 0.99 ($\sim0.8$), but has been manually confirmed to be real.

Briefly, we list the preconditioning parameters used. The standard continuum subtraction offered by \textsc{SoFiA 2} is used with \texttt{contsub.enable = True}. To correct for spectral noise variations across the cube, we use \texttt{scaleNoise.enable = True} together with \texttt{scaleNoise.mode = spectral}. To correct for spatial ripple effects, we enable the ripple filter with \texttt{rippleFilter.enable = True}.

Lastly, we implement manual cross-matching with DES \textit{i,r,g}-bands to remove false positives that were missed by the reliability filter. The majority of galaxies removed in this step were near bright continuum sources and are likely the result of imperfect continuum subtraction and/or variable point spread function dimensions.

After these steps, we are left with a \textsc{SoFiA 2} generated catalogue of 35 \hi sources.

\subsection{Grouping of galaxies}
\label{sec:grouping}

When grouping galaxies, there are two main approaches: Halo-based finders such as in \citep[e.g.][]{Yang_2005, Yang_2007, Yang_2021, Lim_2017} or "Friends-of-Friends" algorithm-based finders \citep[e.g.][]{Berlind_2006, Crook_2007, Tempel_2014}. "Friends-of-Friends" associates different galaxies with one another using two linking lengths that are either fixed or scaled with the redshift of a galaxy. The linking lengths $R_L$ and $V_L$ define lengths in projection and velocity, respectively, by which to group galaxies \citep{Huchra_1982}. Linking lengths are often difficult to choose, as one set of parameters does not satisfy all relevant tests, based on the multiplicity of groups, projected size, and velocity dispersion distributions \citep[see the appendix in][]{Berlind_2006}. It is then common practice to choose linking parameters that satisfy the specific science goals at hand, leading to a large variation in the commonly used linking lengths.

In this work, we implement a simple "Friends-of-Friends" algorithm similar to \citet{Crook_2007} in an analysis of the Two Micron All-Sky Survey (2MASS). The 2MASS redshift survey (2MRS) covers $91\,\%$ of the sky, is $97.6\,\%$ complete at $\textrm{K}_s\leq11.75$~mag, and extends to a depth of $z=0.05$ \citep{Huchra_2012}. In \citet{Crook_2007} $R_L$ is scaled to account for bias introduced by the variation in sampling the luminosity function with redshift while $V_L$ is fixed. We choose a naive prescription where $R_L$ and $V_L$ are fixed parameters. The values chosen for this work are $R_L = 1.2\,\mathrm{Mpc}$ and $V_L = 500\,\rm{km}\,\rm{s}^{-1}$. The value of $V_L$ is similar to $V_0$ in \citet{Ramella_1997} in an analysis of the northern CfA redshift survey, while the value of $R_L$ is between $0.89$ in \cite{Ramella_1997} and $1.6$ in \citet{Crook_2007}. Both parameters fall well within the reasonable range investigated in \citet{Crook_2007}. The science goals of this paper are not to establish ideal linking parameters for \hi group finding, but rather to probe the environmental effects on galaxies in high-density environments such as the one reported here. Choices of $R_L$ between 1$\,\mathrm{Mpc}$ and 1.6$\,\mathrm{Mpc}$ and choices of $V_L$ between 350\kms and 600\kms result in a small variation in the number of galaxies included in the structure, but the structure itself maintains its projected filamentary appearance and the primary scientific results are not impacted.

\begin{figure}
	\includegraphics[width=\columnwidth]{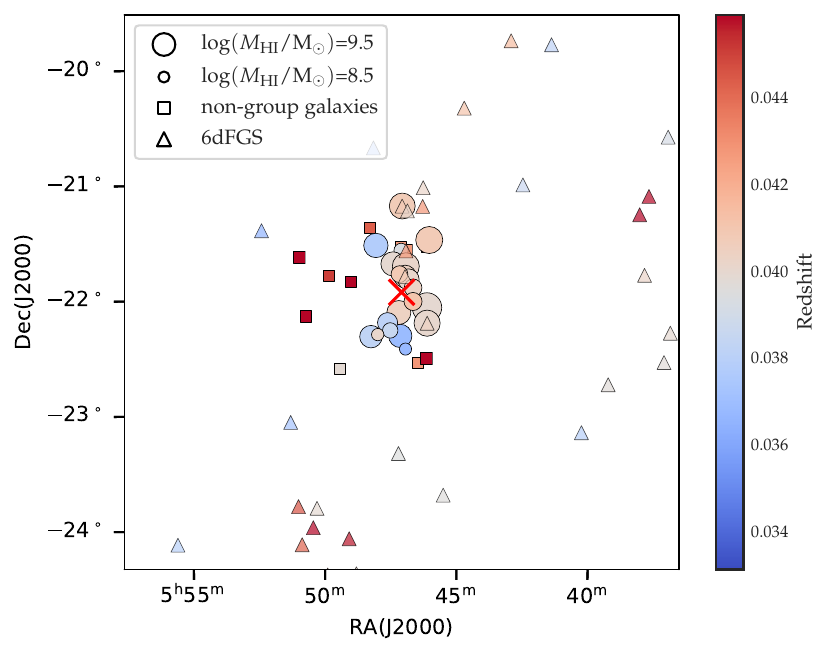}
    \caption{Spatial distribution scatter plot in RA and Dec of all galaxies in the final \textsc{SoFiA} 2 catalogue. Circular markers have been linked by a "Friends-of-Friends" algorithm described in Sec.~\ref{sec:grouping}. Square markers indicate isolated galaxies or group galaxies belonging to groups with fewer members than the primary group. The size of the circular markers is scaled by \hi mass with two indicative masses displayed in the legend. The red cross represents the geometric mean coordinates of the orange points. Galaxies from 6dFGS with spectroscopic redshifts are shown with triangles. All markers are colourised by redshift.}
    \label{spatial_dist}
\end{figure}

In Fig.~\ref{spatial_dist}, the distribution of galaxies is shown for all detections produced by our \textsc{SoFiA} 2 strategy. For reference, the geometric mean of the largest group is indicated by a red cross. Of the 35 sources in the \textsc{SoFiA 2} catalogue, 20 are grouped by our chosen linking lengths; these galaxies are shown by the circle markers in Fig.~\ref{spatial_dist}, scaled by \hi mass. Although the remaining 15 galaxies (denoted with squares) are reliably detected, they are predominantly isolated and show no signs of interaction. As such, they are excluded from further analysis. We note that the distribution of the galaxies results in a projection of a filamentary-like structure from north to south. All markers are colourised by redshift.

Additionally, we searched for spectroscopic redshifts of galaxies within $3\sigma$ of the redshift distribution of the largest group. We find numerous galaxies with confirmed, reliable spectroscopic redshifts from the 6dF Galaxy Survey \citep[6dFGS,][]{Jones_2009}. The filamentary structure observed in the \hi identified galaxies in this work is supported by the distribution of spectroscopic redshifts from 6dFGS. With the uniform coverage from 6dFGS extending beyond the observation’s bounding box, the spectroscopic redshifts still reveal a north-to-south structure. The filamentary projection of the group is not due to sensitivity limits, as we detect a sub-Milky Way mass galaxy ($\log(M_{\rm{HI}}/\rm{M}_{\odot})=8.68$) $\sim~44.3$' from the pointing centre.

\subsection{\textsc{H\,i} detections and properties}
\label{sec:hi_detections}

\begin{figure*}
	\includegraphics[width=0.85\textwidth]
    {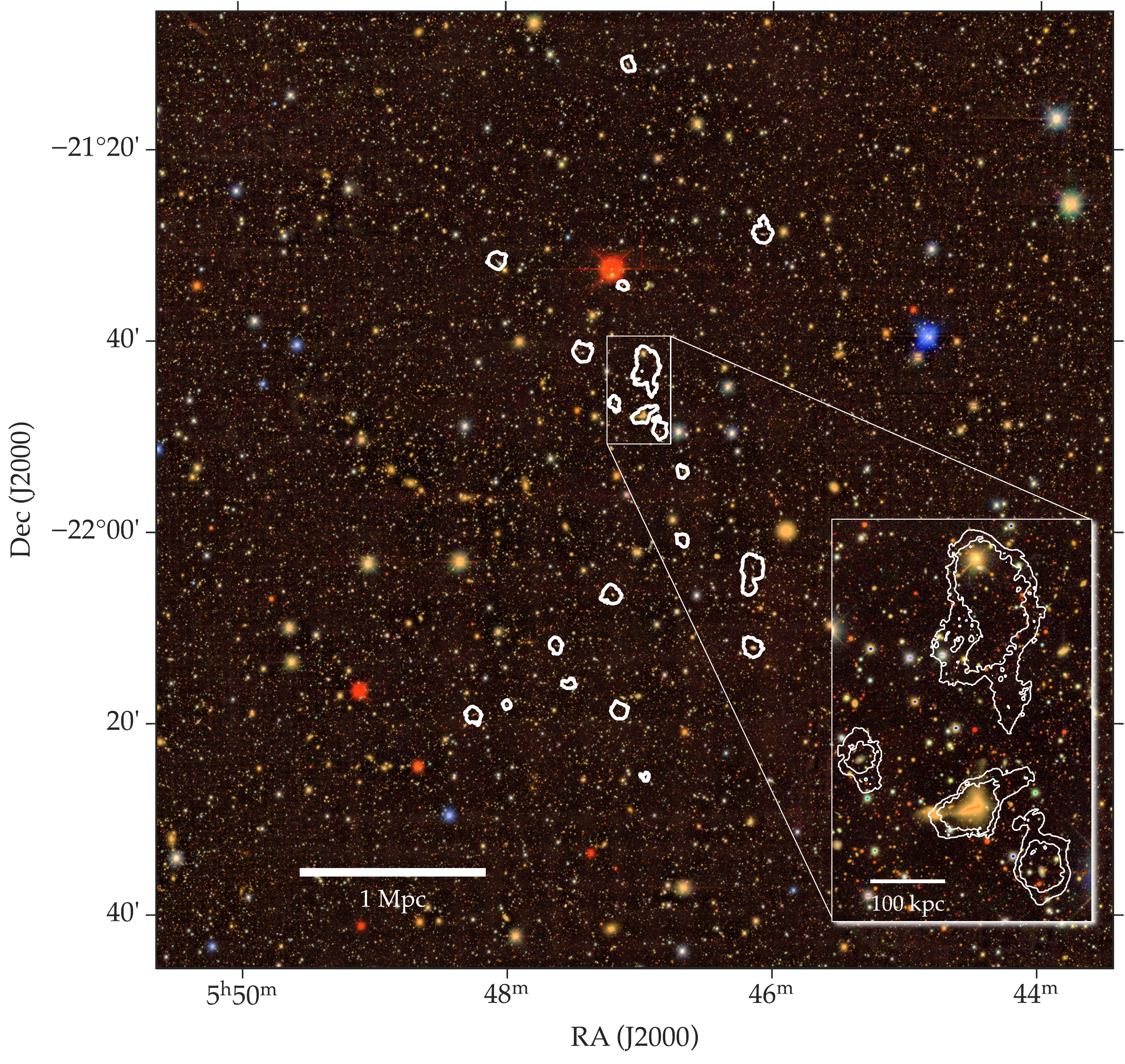}
    \centering
    \caption{DES RGB image constructed from \textit{i}, \textit{r}, and \textit{g} bands. MeerKAT \hi contours overlaid in white at an \hi column density of $1\times\,10^{19}\rm{cm}^{-2}$ at the average galaxy redshift of $z\,=\,0.0395$. A zoomed inset of the compact group discussed in Sec~\ref{sec:comp_group} is shown with white \hi contours at $1\times\,10^{19}\rm{cm}^{-2}$ and $1\times\,10^{20}\rm{cm}^{-2}$.}
    \label{des_image}
\end{figure*}

\begin{landscape}
\begin{figure}
    \centering
    \vspace{0.5cm}
    \includegraphics[height=0.80\textheight]{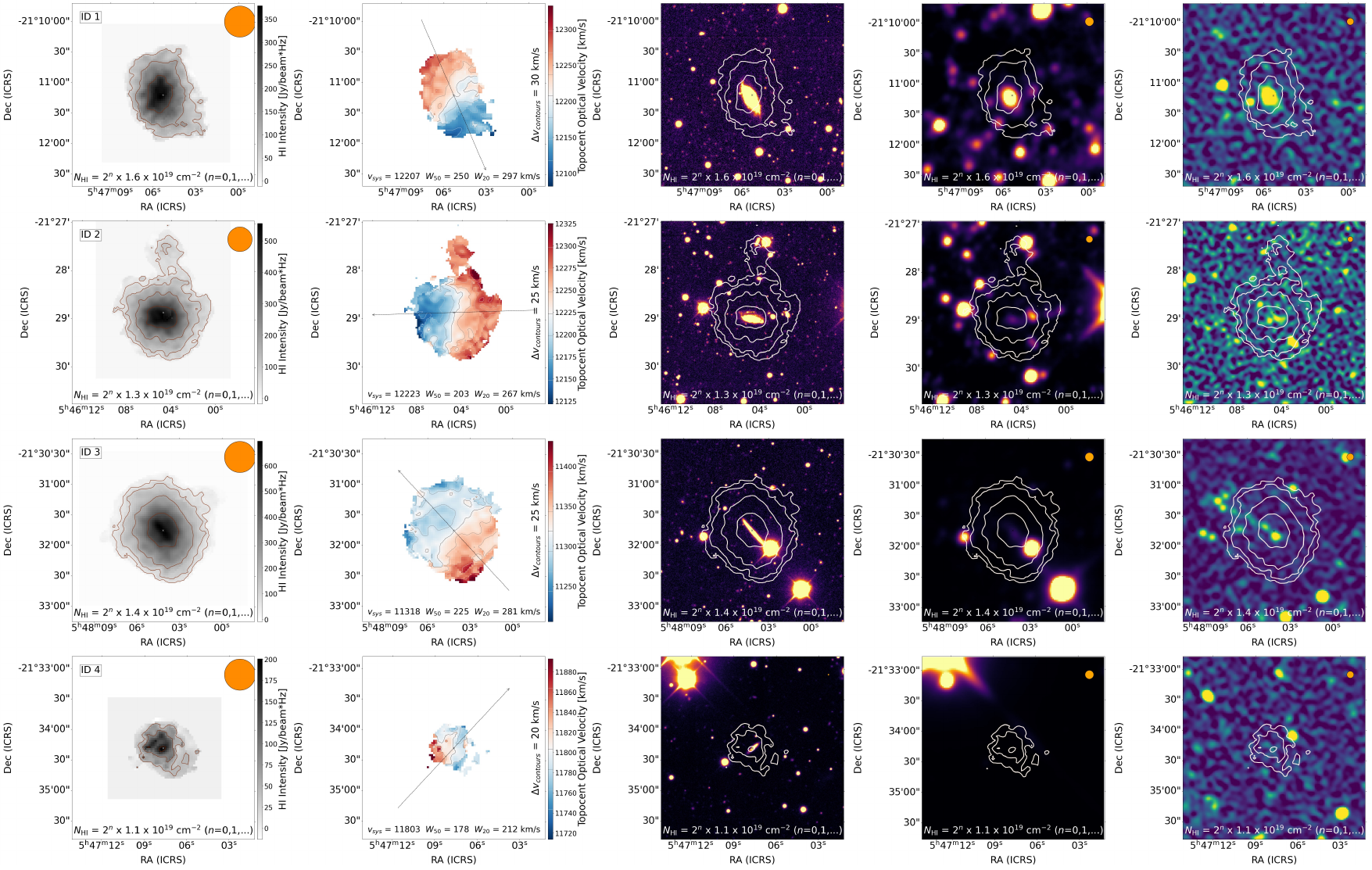}
    \caption{From left to right (for IDs 1-4), the moment 0 (integrated \textsc{H\,i}) map, moment 1 (velocity field), DES \textit{g}-band, WISE W1 band, MeerKAT L-band continuum. In the moment 1 map we show the recessional velocity of the source ($v_{\rm{sys}}$), spectral line width at 50$\,\%$ and 20$\,\%$ of the peak flux density ($W_{50}$ and $W_{20}$), as well as the velocity difference between contours ($\Delta v_{\rm{contours}}$). An indication of the kinematic position angle is given by the line centred on the flux-weighted centroid of the source. In the other images, we show contours from the moment 0 map overlaid at \hi column densities shown at the bottom of the images. The moment 0 map has the \hi restoring beam ($30.65"\times30.65"$) displayed in the top right corner. Similarly, we show the MeerKAT continuum map restoring beam ($6.87"\times6.87"$). The DES \textit{g}-band and WISE W1 band have a point-spread function FWHM of $1.11"$ and $6.1"$ respectively, also shown in the top right.}
    \label{fig:page1}
\end{figure}
\end{landscape}

Focusing only on the galaxies associated with the "Friends-of-Friends" algorithm, we have 20 clear detections with at least 22 galaxies. In Fig.~\ref{des_image} we show a DES \textit{RGB} image constructed from \textit{i},\textit{r}, and \textit{g} bands where the 20 group detections are overlaid with white contours at an \hi column density of $1\times10^{19}\,\rm{cm}^{-2}$ assuming the average redshift, $z=0.0395$. The scale bar in the bottom left is based on the average galaxy redshift. The source properties are listed in Table~\ref{tab:galaxy_properties}, where IDs are ordered by descending declination. Fig.~\ref{fig:page1} shows the moment 0 (integrated \textsc{H\,i} flux), moment 1 (\textsc{H\,i} velocity field), DES \textit{g}-band ($473\,\rm{nm}$ centre wavelength), WISE W1 band ($3.4\,\mu \rm{m}$ centre wavelength), and the MeerKAT L-band (900 - 1670$\,\rm{MHz}$) continuum map for each source in that order from left to right, these plots were made in part with the \textsc{SoFiA image pipeline}\footnote{https://github.com/kmhess/SoFiA-image-pipeline} \citep{Hess_2022}. The integrated spectra for all galaxies and the multiwavelength images of the remaining sources are in appendix~\ref{app:a} and \ref{app:b} respectively. The error bars on the integrated spectra are derived from the per channel RMS of the local noise within the bounding box of the source. The contours overlaid on the moment 0, DES, WISE, and continuum maps correspond to \hi column densities stated at the bottom of the image. \hi column densities are calculated as
\begin{equation}
    N_{\rm{HI}} = 2.33 \times 10^{20} (1+z)^{4} \left( \frac{S}{\rm{JyHz}} \right) \left( \frac{ab}{\rm{arcsec}^{2}} \right)^{-1},
    \label{eq:hi_col_den}
\end{equation}
where $z$ is extracted from the redshift corresponding to the central frequency of each source, $S$ is the integrated \hi flux of the source in $\rm{JyHz}$, and $a$ and $b$ are the major and minor axes of the restoring beam in $\rm{arcsec}$ respectively \citep{Meyer_2017}. The moment 1 map is created using pixels in the moment 0 map, excluding values below the lowest contour. The $v_{\rm{sys}}$, $W_{50}$, $W_{20}$, and $\Delta v$ contours are calculated using the relation between frequency and velocity, that is
\begin{equation}
    V = \rm{c} \frac{\nu_0 - \nu}{\nu},
\end{equation}
where $\nu_0$ is the rest frequency of the spectral line, $\nu$ is the observed frequency, and $\rm{c}$ is the speed of light. $v_{\rm{sys}}$ uses the central frequency of the spectrum of the source, while $W_{50}$ ($W_{20}$) describes the width of the spectral profile where the flux density has increased by more than 50 (20) \% of the peak flux density moving from the outer channels of the cubelet inwards. We adopt a conservative approach to uncertainties associated with the $W_{20}$ and $W_{50}$ by taking an indicative uncertainty of the channel width, $46\,\rm{km}\,\rm{s^{-1}}$. While this represents a relatively large uncertainty when compared to typical galaxy widths, it does not affect the main scientific conclusions of this paper. The line in the moment 1 map indicates the kinematic semi-major axis, with direction pointing to the high-frequency channels. The kinematic semi-major axis of a source is determined by calculating the flux-weighted centroid in each channel where the source has emission. Plotting these points will result in a straight line in the event that the source is a regularly rotating galaxy. Fitting a straight line to these points on the sky using orthogonal regression results in a good approximation of the semi-major axis (see \textsc{SoFiA 2} documentation for details). We note that kinematic analysis for single- or two-channel sources could be inaccurate. This is due to the fairly coarse spectral resolution of $209\,\rm{kHz}$. \hi mass is calculated following the prescription in \citet{Meyer_2017}:
\begin{equation}
    \left( \frac{M_{\rm{HI}}}{\rm{M}_\odot} \right) = 49.7 \left(\frac{D_{\rm{L}}}{\rm{Mpc}} \right)^2 \left( \frac{S}{\rm{JyHz}} \right),
\end{equation}
where $S$ is the integrated \hi flux of the source in $\rm{JyHz}$. $S$ is calculated by summing the flux densities across the source. The \hi redshift calculated from the central frequency of each source is used to calculate the luminosity distance, $D_{\rm{L}}$, of the source. Uncertainties associated with $D_{\rm{L}}$ are derived from \textsc{SoFiA 2} statistical uncertainties in the centroid channel location along the frequency axis. To obtain an estimate for the uncertainty in the integrated flux $S$, we calculate the signal in four emission-free masked regions near the source, for this we use the unique 3D mask output by \textsc{SoFiA 2} for each galaxy. We then take the mean RMS scatter of the flux values as our uncertainty \citep{Ponomareva_2021}. This is then combined in quadrature with a $10\,\%$ absolute flux uncertainty. Uncertainties are shown in Table~\ref{tab:galaxy_properties}.

The SFR and $M_\ast$ for viable sources are estimated in the following way: We manually crossmatch sources using \textsc{CARTA} with DES \textit{i,r,g}-bands, WISE W1, W2, and W3 bands, moment 0 cutouts produced by \textsc{SoFiA}, and the WISE All-Sky Source Catalog. For \hi sources that have been matched to a source in the WISE catalogue, we inspect the \texttt{ext\_flg}. If \texttt{ext\_flg} = 0, the source is consistent with a point source, and we use the largest non-contaminated aperture (checked with flags provided in the WISE catalogue for each aperture) to acquire a magnitude for that band. If the source is extended (\texttt{ext\_flg} $> 0$), we manually ensure that the aperture adequately covers the source. Finally, we inspect the upper limit flag supplied by WISE to determine if the magnitude quoted is a $95\,\%$ upper limit. If the source is a non-detection (as well as no contamination present), we derive an upper limit from the appropriate WISE band sensitivity.

Light emitted from K- and M-type giant stars dominate the W1 and W2 bands, making them good tracers of the stellar mass within a galaxy by tracing the continuum emission from the evolved stellar population of a galaxy \citep{Cluver_2014}. The initial relation in \citet{Cluver_2014} has a large observed scatter, correspondingly derived $M_\ast$ values have large uncertainties. Since then, improvements have been made on these relations \citep{Jarrett_2023}. We choose to use a simplistic relation from \citet{Jarrett_2023}, only considering the W1 magnitude. This is motivated by the sensitivity of the W1 band in WISE, as well as eliminating the uncertainty associated with the ambiguous W2 emission of various sources. The relation is as follows:
\begin{equation}
    \label{eq:stellar}
    \log{M_\ast} = A_0 + A_1\left(\log{L_{\rm{W1}}}\right) + A_2\left(\log{L_{\rm{W1}}}\right)^{2} + A_3\left(\log{L_{\rm{W1}}}\right)^{3}
\end{equation}
with
\begin{equation}
    L_{\rm{W}1}\,(L_\odot) = 10^{-0.4(M-\rm{M_{sun}})}
\end{equation}
where $\rm{M_{sun}} = 3.24$ is the absolute Vega magnitude of the sun, $M$ is the W1 absolute magnitude of the source, and the A coefficients are -12.62185, 5.00155, -0.43857, 0.01593, respectively. $D_{\rm{L}}$, calculated from the \hi redshift, is used to convert the W1 Vega magnitude to an absolute magnitude for each source. Upper limits are indicated in Table~\ref{tab:galaxy_properties}, while contaminated sources or sources where no adequate aperture could be found have their $M_\ast$ values omitted.

WISE W3 band luminosity is dominated by warm dust and polycyclic aromatic hydrocarbons (PAHs) for $\sim L^\ast$ star-forming galaxies, with the contributions being 62.5$\,\%$ and 34$\,\%$ respectively \citep{Cluver_2017}. This makes W3 a tracer of star formation. However, for these galaxies stellar continuum contributes 15.8$\,\%$ of the W3 light. To correct for this effect we subtract 15.8$\,\%$ of the flux in W1 from W3 before using the relation. SFR values for galaxies with corresponding W3 magnitudes are calculated using Eq.~4 in \citet{Cluver_2017}, calibrated using Spitzer Infrared Nearby Galaxy Survey \citep[SINGS;][]{Kennicutt_2003} and Key Insights on
Nearby Galaxies: a Far-Infrared Survey with Herschel \citep[KINGFISH;][]{Kennicutt_2011}:
\begin{equation}
    \log \rm{SFR} (\rm{M}_\odot\, \rm{yr}^{-1}) = (0.889\pm0.018)\log L_{12\mu m} (L_\odot) - ( 7.76\pm0.15),
\end{equation}
where $L_{12\mu m}$ is the luminosity in the WISE W3 band. We obtain flux densities for the W3 and W1 bands by following the outline given in the WISE data processing documentation\footnote{https://wise2.ipac.caltech.edu/docs/release/allsky/expsup/sec4\_4h.html}:
\begin{equation}
\label{eq:vegatoflux}
    {F}_{\nu}(\mathrm{Jy})= \mathrm{F}_{\nu0}\times10^{\left(-\mathrm{m_{Vega}/2.5}\right)},
\end{equation}
where values for $\mathrm{F}_{\nu0}$ are given in the documentation. Similarly to the $M_\ast$ calculation given by Eq.~\ref{eq:stellar}, we omit contaminated sources and indicate WISE quoted 95$\,\%$ upper limits, as well as sensitivity derived upper limits in Table~\ref{tab:galaxy_properties}. Finally, we define $f_{\rm{HI}}^\ast = \log_{10}(M_{\rm{HI}}/M_\ast)$ for sources where stellar masses are calculated.

\begin{table*}
	\centering
	\caption{Properties of group galaxies/sources using \hi and infrared data.}
    \label{tab:galaxy_properties}
    \begin{tabular}{rlllllllll}
    \toprule
    ID & RA (J2000) & Dec (J2000) & $z_{\rm{HI}}$ & $D_{\rm{L}}$ & $\rm{log}$ & $W_{50}$ & $\rm{log}$ & $\rm{SFR}$ & ${f}^{*}_{\rm{HI}}$ \\
     & (h:m:s) & (d:m:s) &  & (\rm{Mpc}) & $(M_{\rm{HI}}/\rm{M_\odot)}$ & (\kms) & $(M_*/\rm{M_\odot})$ & ($\rm{M_\odot\,\rm{yr^{-1}}}$) & \\
    \midrule
    1  & 5:47:05.3  & -21:11:11.89 & 0.0407 & 186.0 $\,\pm\,0.12$ & 9.832 $\,\pm\,0.044$ & 250 & ... & ... & ... \\
    2  & 5:46:04.7  & -21:28:51.63 & 0.0408 & 186.2 $\,\pm\,0.06$ & 9.988 $\,\pm\,0.048$ & 203 & 9.24 & 0.18 & 0.74 \\
    3  & 5:48:04.2  & -21:31:44.92 & 0.0378 & 172.1 $\,\pm\,0.07$ & 9.565 $\,\pm\,0.048$ & 226 & 9.50 & 0.19 & 0.06 \\
    4  & 5:47:07.7  & -21:34:18.03 & 0.0394 & 179.6 $\,\pm\,0.24$ & 8.616 $\,\pm\,0.061$ & 179 & ... & ... & ... \\
    5  & 5:47:25.9  & -21:41:19.86 & 0.0399 & 182.2 $\,\pm\,0.05$ & 9.593 $\,\pm\,0.048$ & 171 & ... & ... & ... \\
    6  & 5:46:57.1  & -21:42:33.99 & 0.0401 & 183.2 $\,\pm\,0.03$ & 9.965 $\,\pm\,0.055$ & 152 & ... & ... & ... \\
    7  & 5:47:11.2  & -21:46:40.14 & 0.0408 & 186.4 $\,\pm\,0.10$ & 8.843 $\,\pm\,0.045$ & 111 & <8.67 & <0.52 & >0.17 \\
    8  & 5:46:58.6  & -21:47:54.69 & 0.0405 & 184.7 $\,\pm\,0.19$ & 9.438 $\,\pm\,0.059$ & 542 & 10.98 & 3.27 & -1.54 \\
    9  & 5:46:51.0  & -21:49:27.23 & 0.0401 & 182.9 $\,\pm\,0.08$ & 9.266 $\,\pm\,0.053$ & 138 & 9.27 & 0.21 & -0.00 \\
    10 & 5:46:40.6  & -21:53:47.16 & 0.0403 & 183.9 $\,\pm\,0.10$ & 8.916 $\,\pm\,0.055$ & 137 & 8.56 & <0.12 & 0.36 \\
    11 & 5:46:40.7  & -22:00:53.58 & 0.0411 & 187.8 $\,\pm\,0.09$ & 8.964 $\,\pm\,0.057$ & 91  & <8.68 & <0.53 & >0.28 \\
    12 & 5:46:09.3  & -22:04:00.44 & 0.0399 & 182.3 $\,\pm\,0.02$ & 10.314 $\,\pm\,0.045$ & 269 & 9.61 & 0.56 & 0.70 \\
    13 & 5:47:12.5  & -22:06:39.34 & 0.0404 & 184.4 $\,\pm\,0.06$ & 9.584 $\,\pm\,0.043$ & 175 & 9.22 & 0.20 & 0.37 \\
    14 & 5:47:37.9  & -22:11:56.29 & 0.0382 & 173.9 $\,\pm\,0.09$ & 9.096 $\,\pm\,0.063$ & 169 & 8.69 & <0.07 & 0.40 \\
    15 & 5:46:08.9  & -22:12:09.80 & 0.0400 & 182.8 $\,\pm\,0.05$ & 9.852 $\,\pm\,0.046$ & 229 & 10.16 & 2.14 & -0.30 \\
    16 & 5:47:31.8  & -22:15:59.01 & 0.0386 & 175.8 $\,\pm\,0.16$ & 8.682 $\,\pm\,0.058$ & 76  & <8.46 & <0.04 & >0.22 \\
    17 & 5:48:00.0  & -22:18:09.32 & 0.0401 & 183.0 $\,\pm\,0.10$ & 8.498 $\,\pm\,0.060$ & 63  & <8.66 & <0.50 & >-0.16 \\
    18 & 5:47:09.2  & -22:18:50.47 & 0.0368 & 167.7 $\,\pm\,0.06$ & 9.415 $\,\pm\,0.047$ & 194 & <8.58 & <0.43 & >0.83 \\
    19 & 5:48:15.2  & -22:19:13.58 & 0.0382 & 174.3 $\,\pm\,0.07$ & 9.358 $\,\pm\,0.058$ & 169 & ... & ... & ... \\
    20 & 5:46:57.3  & -22:25:41.12 & 0.0367 & 167.3 $\,\pm\,0.17$ & 8.410 $\,\pm\,0.059$ & 58  & ... & ... & ... \\
    \bottomrule
    \vspace{2mm}
    \end{tabular}
    \begin{minipage}{\textwidth}
    Cols (1)-(3): Source IDs sorted by decreasing declination, RA and Dec from \hi flux weighted centroid produced by \textsc{SoFiA 2}. Cols (4)-(5): \hi derived redshift and luminosity distance with uncertainties. Cols (6)-(7): \hi mass with uncertainties and $W_{50}$ (uncertainty of $46\,\rm{km}\,\rm{s^{-1}}$) of integrated \hi line profile outlined in~Sec.~\ref{sec:hi_detections}. Cols (8)-(10): Stellar mass (uncertainty of $0.12\,\rm{dex}$), star formation rate (uncertainty of $0.15\,\rm{dex}$), and \hi gas fraction, described in Sec.~\ref{sec:hi_detections}.
    \end{minipage}
\end{table*}

\section{Discussion}
\label{sec:discussion}

Sources within the structure are classified into "disk-like", "interacting system", or "unclear" based on visual inspection of the DES imaging of the optical counterpart to the \hi detection. We find IDs 1, 3, 5, 13, 14, 15, and 19 to be disk-like galaxies. Interacting systems are apparent in IDs 2, 6, 8, and 12. For IDs 4, 7, 9, 10, 11, 16, 17, 18, and 20 their morphologies are not clear based on the data inspected. Inspection of the \hi spectra of the sources finds double horn profiles for IDs 1, 4, 6, 9, 12, 14, 15, 18, and 19, though some appear to be asymmetric. Note that this approach only loosely classifies the sources, as we acknowledge the subjective nature of the methodology.

\begin{figure}
	\includegraphics[width=\columnwidth]{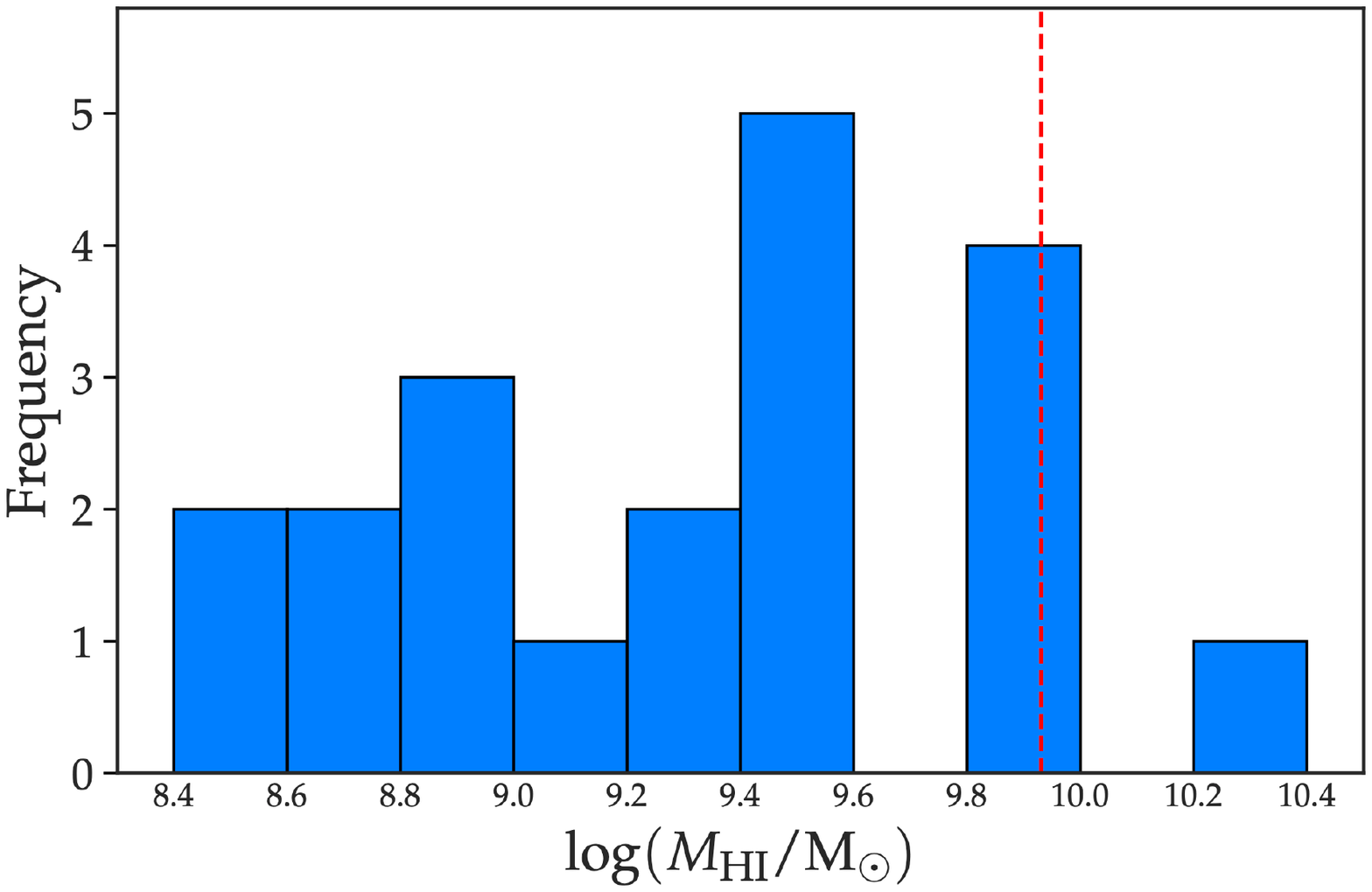}
    \caption{\hi mass distribution of galaxies in bins of $0.2\,\rm{dex}$. The vertical dashed red line indicates the "knee", $M_*$, of the Schechter function fit to the HIMF at $z=0$ \citep{Jones_2018}.}
    \label{fig:hi_mass_distribution}
\end{figure}

The \hi mass distribution of the grouped galaxies sample is shown in Fig.~\ref{fig:hi_mass_distribution} as a histogram. The masses are divided into 10 bins of equal size ($\sim 0.2\,\rm{dex}$) spanning the range of $8.5 \leq \log(M_{\rm{HI}}/\rm{M}_\odot) \leq 10.5$. All 20 sources fall within $\Delta \log(\frac{M_{\rm{HI}}}{\rm{M}_\odot}) = 1.9$. A comparison to the $z=0$ HIMF based on Arecibo Legacy Fast ALFA (ALFALFA) blind \hi survey of the nearby universe provides valuable insight into the sample in question. The Schechter function fit has a "knee" mass of $m_\ast = 9.94 \pm 0.01$ \citep{Jones_2018}. ID 2, ID 6, and ID 12 all have \hi masses above the "knee" and are discussed in depth in Sec.~\ref{sec:interactions}.

\subsection{Scaling relations}
\label{sec:scaling_relations}

\begin{figure}
    \includegraphics[width=\columnwidth]{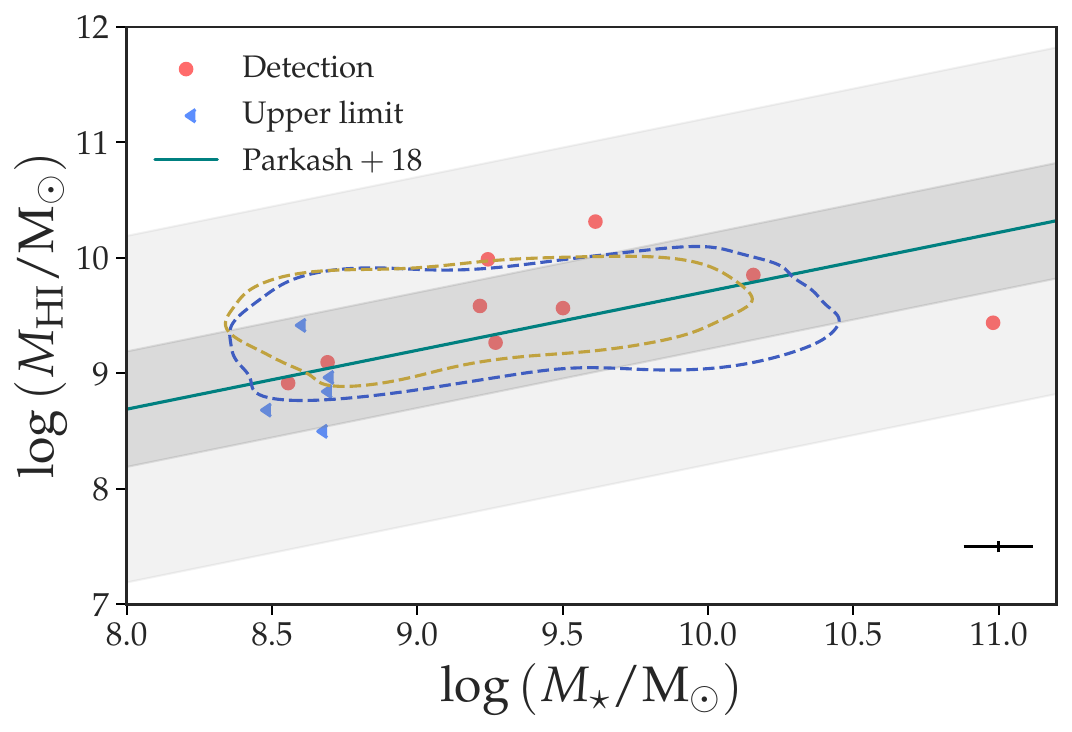}
    \caption{$M_{\rm{HI}} \text{--} M_\ast$ relation from \citet{Parkash_2018} in teal, with $1\sigma$ and $3\sigma$ confidence intervals shown with the shaded areas. Sources from this work are shown in coral (detections) and blue (upper limits). An indicative error is given in the bottom right as $0.12\,\rm{dex}$ on the stellar mass (based on the scatter of the relation \citep{Jarrett_2023}) and $0.05\,\rm{dex}$ on the \hi mass (based on the uncertainty outlined in Section~\ref{sec:hi_detections}). The blue (yellow) dashed contours enclose 68$\,\%$ of \textsc{Simba} analogue (field) galaxies; see Sec.~\ref{subsec:simba_sr}.}
    \label{fig:hi_vs_sm}
\end{figure}

We investigate the $M_{\rm{HI}} \text{--} M_\ast $ relation with a comparison of linked galaxies within the group structure to the stellar-\hi mass relation for \textsc{H\,i}-selected galaxies reported in \citet{Parkash_2018}, see Fig.~\ref{fig:hi_vs_sm}. Of the 9 sources with clear $M_\ast$ values, we find that 6 (3) fall within the $1\sigma$ ($3\sigma)$ confidence interval. All but one \hi detections with stellar mass upper limits (non-detections or upper limits quoted by WISE) are within $1\sigma$ of the relation. The most notable outlier is ID 8 with the highest stellar mass of $\log (M_\ast/\rm{M}_\odot) = 10.98$; see Sec.~\ref{sec:comp_group} for an expanded discussion on this system. In general, we find the galaxies to be in good agreement with \citet{Parkash_2018}. We note that IDs 1, 4, 5, 6, 19, and 20 are all either quoted as contaminated by the WISE catalogue or clearly contaminated as per a visual inspection. These \hi galaxies all have counterparts in the DES imaging, and have hints of, mostly contaminated, counterparts in the WISE bands or catalogue.

\begin{figure}
	\includegraphics[width=\columnwidth]{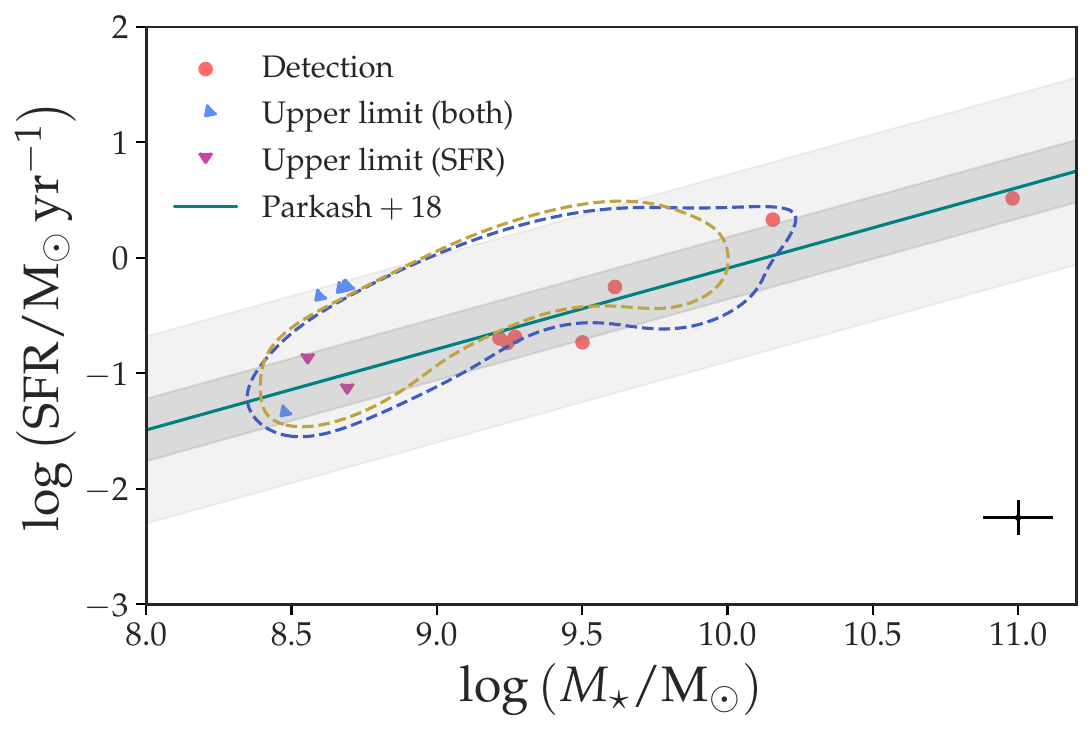}
    \caption{SFMS ($\rm{SFR} \text{--} M_\ast$ relation) from \citet{Parkash_2018} in teal, with $1\sigma$ and $3\sigma$ confidence intervals shown with the shaded areas. Sources from this work are shown in coral (Detections), purple (upper limits on SFR), and blue (upper limits on SFR and stellar mass). An indicative error on the stellar mass and SFR is given in the bottom right as $0.12\,\rm{dex}$ and $0.15\,\rm{dex}$ respectively based on the scatter of the relations \citep{Jarrett_2023, Cluver_2017}. The blue (yellow) dashed contours enclose 68$\,\%$ of \textsc{Simba} analogue (field) galaxies, see Sec.~\ref{subsec:simba_sr}.}
    \label{fig:sfr_vs_sm}
\end{figure}

The relationship between SFR and $M_*$ is also investigated with a comparison to the \citet{Parkash_2018} relation. The star-forming main sequence (SFMS) compared to in Fig.~\ref{fig:sfr_vs_sm} is based on a spiral sample. Of the detections, 5 are within the $1\sigma$ confidence interval from the relation, and 2 are outside the $1\sigma$ level but within the $3\sigma$ confidence interval. For SFR upper limits, both are within $1\sigma$ of the relation. For SFR and $M_*$ upper limits, one is within $1\sigma$ and 4 are within $3\sigma$. ID 8 has the highest SFR ($(\rm{SFR}/\rm{M_\odot\,\rm{yr^{-1}}}) = 3.27$) and is discussed in Sec.~\ref{sec:comp_group}.

We caution against over-interpretation of the comparisons with \citet{Parkash_2018} due to differences in selection criteria \citep[see][]{Parkash_2018}. However, we wish to see if there are any clear correlation differences between an \textsc{H\,i}-selected sample and field galaxies despite the selection differences. We conclude that the group galaxies identified in this work are consistent with what we expect from local star forming galaxies both in the SFMS and the $M_{\rm{HI}}-M_*$ relations.

\subsection{Compact group}
\label{sec:comp_group}

Of the most interesting interactions is a compact group of 4 sources, highlighted in Fig.~\ref{des_image}, found near the geometric centre of the structure. Source IDs 6, 7, 8, and 9. The central frequency of the compact group corresponds to $z= 0.0405$, and a luminosity distance of $184.3 \,\rm{Mpc}$. The centre channels of these sources fall within $\sim210$\kms of each other. A rectangular bounding box of the four sources is $\sim 300 \times 500 \,\rm{kpc}$. ID 6 shows the clearest signs of interaction, with a large emission tail extending to the south. The compressed contours to the north and the extended \hi emission to the south may be indicative of ram-pressure stripping. However, the intragroup medium is typically less dense than the ICM thought to be the primary cause of ram-pressure stripping seen in clusters \citep{Gunn_1972}. Additionally, the close proximity to IDs 7, 8, and 9 suggests tidal stripping. This tail is shown at an \hi column density of $1.5 \times 10^{19} \,\rm{cm}^{-2}$ in Fig.~\ref{fig:page2}. A foreground star contaminates all WISE bands, and no NIR emission can confidently be associated with the \hi galaxy. The optical host galaxy can be seen in the DES \textit{g}-band directly south of the aforementioned foreground star. ID 7 seems to be only mildly affected by the neighbouring galaxies, with a disturbed \hi rotation visible in the moment 1 map, and an asymmetrical spectral profile. ID 7 also has a subtle elongation in its lowest \hi contour to the southwest. Similarly to ID 6, ID 8 appears to be an influential factor in the interactions occurring. ID 8 consists of 2 galaxies, both of which can be identified in the DES \textit{g}-band image. While the moment 1 map does not show abnormal rotation, the spectral profile has the largest extent ($W_{50} = 541$\kms) of all sources in the full sample and represents a bimodal Gaussian distribution characteristic of two interacting galaxies. We see a clear extended continuum counterpart aligned with the highest density \hi contour, potentially indicating AGN activity or disk star formation. Finally, there is a faint \textsc{H\,i}-tail extending to the northwest. A single entry in the WISE catalogue allows us to calculate the stellar mass and SFR, from this ID 8 is found to have the highest stellar mass at $\log (M_\ast/\rm{M}_\odot) = 10.98 $ with the caveat that WISE identifies only one galaxy when two are present. The high SFR seen in ID 8 is a common characteristic of similar interacting systems \citep{Laurikainen_1989}. Taking all into consideration, ID 8 is proposed to be an ongoing merging system between two galaxies with a high star formation rate, high stellar mass and potential AGN activity, while the \hi mass of the system is low in comparison. Note that the galaxy east of the interacting pair visible in the DES \textit{g}-band has a spectroscopic redshift of $z \sim 0.048$, and is thus unrelated to the interaction. Lastly, ID 9 has small \hi clouds extending to the north of the galaxy. The rotation of ID 9, shown in the moment 1 map, seems disturbed. ID 9 does have a WISE counterpart and therefore has corresponding SFR and stellar mass estimates. Some continuum emission seems to be present in the vicinity of the \hi galaxy but it is not clear if it is associated. These 4 galaxies appear to be a compact group where the \hi extent of the galaxies is proving to be useful for probing the interactions between these galaxies. We propose that ID 6 is potentially an example of a galaxy that has undergone tidal interactions with the other galaxies in the compact group, before having its gravitationally-disrupted and rarefied, \hi gas ram-pressure stripped by the medium present within the compact group. A combination of tidal interactions and ram-pressure stripping has been studied in detail in the Fornax cluster \citep{Serra_2023}. Given the spectral and spatial resolution of this data, a deeper MeerKAT image in 32k mode would provide a better column density sensitivity to potentially find \hi clouds not detected in this work, as well as provide superior velocity fields through finer spectral resolution, allowing more accurate kinematic analysis. Furthermore, a detailed X-ray study of the medium present in this compact group is required to definitively conclude that the galaxy has undergone ram-pressure stripping.

\subsection{Individual sources of interest}
\label{sec:interactions}

On the northern outskirts of the large-scale structure, we find an interacting system of two galaxies seen in source ID 2. ID 2 consists of a large disk-like galaxy in the south and a smaller disk-like galaxy in the north. The larger southern galaxy has a position angle nearly perpendicular to that of the northern galaxy. ID 2 has a relatively low stellar mass $(\log(M_*/\rm{M}_\odot)=9.24)$ compared to its \hi mass $(\log(M_{\rm{HI}}/\rm{M}_\odot)=9.988)$. This is somewhat expected as the northern galaxy does not appear in the WISE catalogue and so stellar mass and SFR are calculated based solely on the larger southern galaxy, while the \hi mass takes both galaxies into account. For the aforementioned reasons, ID 2 holds the highest non-upper limit neutral gas fraction in the sample.

In source ID 12, we see an analogue of source ID 2, albeit with the primary galaxy north of the satellite galaxy. In ID 12 the inclination angles of the galaxies are not as perpendicular as was the case in ID 2. This is the most \hi massive source in the sample $(\log(M_{\rm{HI}}/\rm{M}_\odot)=10.314)$, with the caveat that the satellite galaxy also contributes to the \hi mass calculation. Similarly to ID 2, we estimate SFR and stellar mass using only the WISE entry for the main galaxy, as the infalling galaxy does not appear in the catalogue. ID 12 also has a particularly high neutral gas fraction that is expected, considering a significant amount of stellar mass is not being accounted for in our estimate. This is perhaps a system that is in the very early stages of merging as the velocity fields are relatively undisturbed, while a clear \hi bridge is visible.

Source ID 15 is in and of itself an interesting spiral galaxy with hints of spiral arms visible in the DES \textit{g}-band image. The galaxy has a high SFR and therefore lies within the upper $3\sigma$ confidence interval of the SFMS. The radio continuum map shows bright continuum emission associated with the galaxy, coincident with the WISE W1, DES \textit{g}-band, and \hi flux weighted centroid, potentially indicating the presence of AGN activity. However, the WISE colour properties (W1-W2 $\sim0.1$ mag and W2-W3 $\sim3.7$ mag) are more consistent with a starburst galaxy than AGN activity \cite{Wright_2010}. The integrated \hi spectrum is an undisturbed double-horn profile. 

\subsection{Group properties}
\label{sec:group_prop}
The \hi grouping consists of 20 \hi detections, all of which have clear DES counterparts. The structure has a spatial extent from the northernmost to the southernmost galaxy of $\sim 3.7\,\rm{Mpc}$. This is calculated using the average group galaxy redshift of $z = 0.04$. Along the line-of-sight, the group extends $20.5\,\rm{Mpc}$ assuming only Hubble flow, with the nearest galaxy at $167.3\,\rm{Mpc}$ and the furthest at $187.8\,\rm{Mpc}$. Clearly, this is too large for the galaxies to be gravitationally bound, however, at the redshift of $z\sim0.04$, peculiar velocities dominate, making the line-of-sight distance uncertain and likely an upper limit. Summing the \hi mass of all the group galaxies, we find a total \hi mass of $\log (M_{\rm{HI}}^{\rm{tot}}/\rm{M}_\odot) = 10.9$, while the average \hi mass of a galaxy is found to be $\log (M_{\rm{HI}}/\rm{M}_\odot) = 9.599\,\pm\,0.053$.

An interesting comparison can be made with a similar \hi group discovery in \citet{Ranchod_2021}. This serendipitous discovery of a group of 20 \hi galaxies is at a similar redshift of $z \sim 0.044$. While more tightly bound along line-of-sight, the group in \citet{Ranchod_2021} is also spatially extended, at $\sim~3\, \rm{Mpc}$. The average galaxy \hi mass is higher at $\log(M_{\rm{HI}}/\rm{M}_\odot) = 9.90$. Total \hi mass is given as $\log (M_{\rm{HI}}^{\rm{tot}}/\rm{M}_\odot) = 11.20$, making the group significantly more \hi massive.

More recently, \citet{Glowacki_2024} has reported the discovery of 49 \textsc{H\,i}-rich galaxies in a 2.3-hour MeerKAT observation. This discovery consists of at least 3 galaxy groups, with multiple sub-groups. Due to significant differences in grouping strategies, we do not make group property comparisons here; we do however highlight the redshifts of the three groups at $z\sim$ 0.033, 0.041, and 0.055.

Although the groups have clear differences, this growing number of discoveries demonstrates the ability of MeerKAT to detect loosely bound \hi groups within a redshift window around $z\sim0.04$.

\subsection{Large-scale structure context with \textsc{Simba}}
\label{sec:lss}

To place the group structure in context within the evolution of the large-scale structure of the universe, we investigate distributions of galaxies using the cosmological hydrodynamical simulation \textsc{Simba} \citep{Dave_2019}, the successor to \textsc{Mufasa} \citep{Dave_2016}.

We intend to provide an indicative sense of how common or extreme our group is. We use the catalogues from the flagship \textsc{Simba}\footnote{http://simba.roe.ac.uk/} run, which has $1024^{3}$ gas particles and $1024^{3}$ dark matter particles within a comoving volume of $100 h^{-1} \,\rm{Mpc}$. The simulation assumes \citet{Planck_2016} cosmological parameters and runs starting from $z=249$ to $z=0$. We query the catalogues on \textsc{ilifu}\footnote{https://www.ilifu.ac.za} using the \textsc{CAESAR}\footnote{https://caesar.readthedocs.io/en/latest/catalog.html} package.

\subsubsection{Detection criteria}
\label{criteria}

We apply individual galaxy detection criteria on the peak \hi flux of the galaxies within the cone. For each galaxy we assume a boxcar spectral profile and calculate the peak flux of this profile, in the following way: \textsc{Simba} provides intrinsic \hi and stellar masses, $M_{\rm{HI}}$ and $M_{\ast}$ respectively, which can be used to obtain an approximate total baryonic mass $M_{\rm{bar}}$ \citep[see][]{Lelli_2019},
\begin{equation}
    M_{\rm{bar}} = 1.33M_{\rm{HI}} + M_{\ast}.
\end{equation}
From the total baryonic mass, we use the Baryonic Tully-Fisher relation \citep{Lelli_2019}, to estimate the mean line-width of the \hi profile at 20$\,\%$ of the peak flux density, $W_{\rm{P20}}$,
\begin{equation}
    \log\left(\frac{M_{\rm{bar}}}{\rm{M}_{\odot}}\right)=(3.75\pm0.08)\log\left(\frac{W_{\rm{P20}}/2}{\rm{km}\,\rm{s}^{-1}}\right)+(1.99\pm0.18).
\end{equation}
We use equation 46 from \citet{Meyer_2017} to obtain rest-frame velocity integrated flux using the intrinsic \hi mass for each \textsc{Simba} galaxy,
\begin{equation}
    \left(\frac{M_{\rm{HI}}}{\rm{M}_\odot}\right)=\frac{2.35\times10^5}{1+z}\left(\frac{D_{\rm{L}}}{\rm{Mpc}}\right)^2\left(\frac{S^{V_{\rm{rest}}}}{\rm{Jy}\,\rm{km}\,\rm{s}^{-1}}\right).
\end{equation}
Under the boxcar approximation for the \textsc{H\,i}-velocity profile, the rest-frame velocity integrated flux becomes
\begin{equation}
    S^{V_{\rm{rest}}}=S_{\nu,\rm{peak}}\Delta V_{\rm{rest}},
    \label{eq:boxcar}
\end{equation}
where $\Delta V_{\rm{rest}}$ is the width, and $S_{\nu,\rm{peak}}$ is the peak of the rest-frame boxcar spectral profile. 

Assuming that $\Delta V_{\rm{rest}}=W_{\rm{P20}}$, and with $W_{\rm{P20}}$ and $S^{V_{\rm{rest}}}$ in hand for each galaxy, we utilise Eq.~\ref{eq:boxcar} to find the peak of the boxcar profile. The detection criterion is chosen to be $S_{\nu,\rm{peak}}>5\,\sigma_{\rm{RMS}}$ where $\sigma_{\rm{RMS}}$ is an empirically determined RMS from our MeerKAT observation. We require that each galaxy is at least 3 channels wide, i.e. $\Delta V_{\rm{rest}}>3\times46$\kms. A qualitative comparison is made to a similar future SKA-mid observation. Here we aim only to show the potential enhancement for galaxy group detection given an SKA-mid observation of the same on-source integration time as our observations ($\sim14$ hours). We scale the RMS from our observation with the factor improvement gained by accounting for the increase in collecting area from adding 80$\times15\,\rm{m}$ diameter SKA-mid dishes to the array.

\subsubsection{Analogue criteria}
\label{analogue_criteria}

To qualitatively illustrate how likely MeerKAT is to discover similar structures we define analogous structures as groupings that meet the following requirements: $10\,\le\,\rm{galaxy}\,\rm{ members}\,\le\,30$, $10.6\,\le\,\log (M_{\rm{HI}}^{\rm{tot}}/\rm{M_\odot})\,\le\,11.4$, and $1.7\,\le\,D_{\rm{p}}\,\le\,5.7 \, \rm{Mpc}$, where the projected extent, $D_{\rm{p}}$, is the maximum projected distance between any two group members in $\rm{Mpc}$. Small variations in these selection criteria do not lead to significant changes in our findings. We then run the same "Friends-of-Friends" algorithm used to associate galaxies in the \textsc{Simba} lightcones, see Sec.~\ref{sec:grouping}. The same linking lengths of $R_L = 1.2\,\mathrm{Mpc}$ and $V_L = 500$ are chosen for consistency.

\subsubsection{Lightcone generation}

To simulate a MeerKAT observation, we construct lightcones from \textsc{Simba} snapshots out to $z=0.1$ following the approach taken in \citet{Lovell_2021}, just below the RFI corrupted region ($0.1\lesssim z \lesssim 0.3$). This is done by combining every second snapshot with a random orientation permutation to avoid repeating large-scale structure. The opening aperture used in the first \textsc{Simba} box is taken to be $\sim1.0^{\circ} \times 1.0^{\circ}$ at a random location on the face of the \textsc{Simba} box. The subsequent apertures are scaled by the length of the lightcone appropriately, while the location of the intersection remains randomised. A catalogue of \textsc{Simba} galaxies within the cone is then created.

\begin{figure*}
    \centering
	\includegraphics[width=1.5\columnwidth]{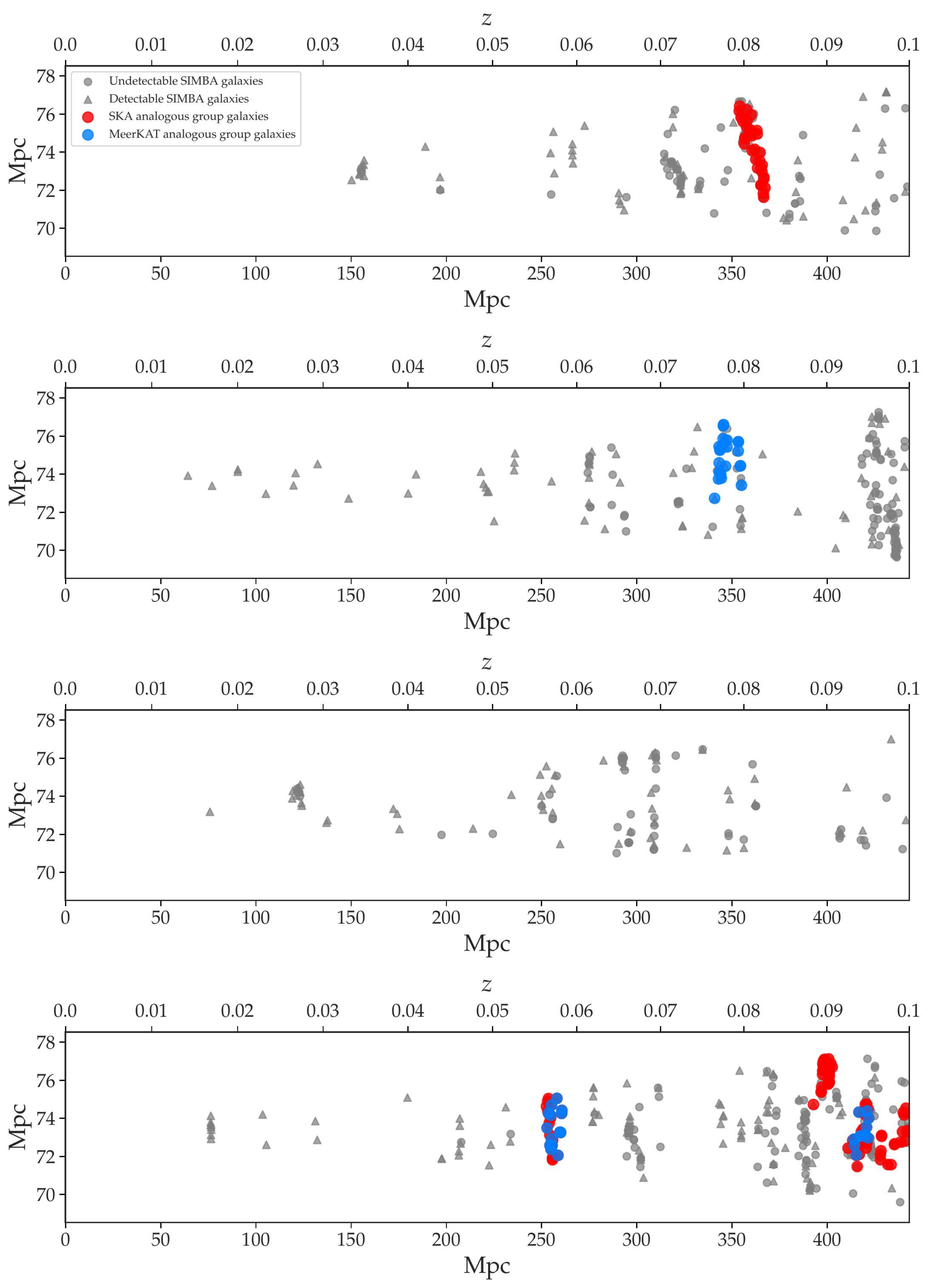}
    \caption{Depiction of example lightcones generated with \textsc{Simba}. The bottom x-axis and the y-axis refer to the cosmological comoving distance. The top x-axis shows the corresponding redshift. Blue points are galaxies that would be detectable with a MeerKAT observation analogous to the one in this work. Red points show galaxies that would be above the detection threshold given a similar observation with the addition of 80 SKA-mid dishes. Grey circles show galaxies that do not form part of either MeerKAT or SKA-mid detectable structures, given an observation with equal integration time. Grey triangles show galaxies that are individually detectable by MeerKAT, but do not belong to a detectable analogues structure.}
    \label{fig:lightcones}
\end{figure*}

In Fig.~\ref{fig:lightcones} \textsc{Simba} lightcones are shown with co-moving cosmological distance in Mpc given on the bottom x-axis and the y-axis and the corresponding redshift on the top x-axis. The grey circles represent all \textsc{Simba} galaxies present within the \textsc{Simba} lightcone. Grey triangles indicate galaxies that are individually detectable by MeerKAT, but do not form part of a detectable analogous structure. The blue points show galaxies that are both detectable by MeerKAT given the aforementioned detection criteria and grouped by the FoF algorithm into structures that match our analogue criteria. The red points are analogous groups that would be detectable given the improved collecting area of SKA-mid, but with the same integration time. In Fig.~\ref{fig:lightcones} we show four sample lightcones that occurred within our simulated observations. In the topmost lightcone, we see an example of a scenario where only SKA-mid would be able to detect the analogous structure at $z\sim 0.08$. In the second example, MeerKAT (and SKA-mid) would be able to detect the analogue at $z \sim0.077$. In the third example, we see numerous galaxies individually detectable by MeerKAT. However, neither MeerKAT nor SKA-mid detect any analogous structures. In the final example MeerKAT detects two analogues while SKA-mid improves these detections by adding members below the MeerKAT detection threshold, as well as detecting an additional analogous structure. These examples showcase that some groups may remain invisible to our instruments as a result of flux sensitivity. However, the analogue criteria also play a role as there are clearly detectable overdensities that simply do not meet the analogue criteria set in this work, highlighting the importance of group structure definitions.

\begin{figure*}
    \centering
	\includegraphics[width=1.6\columnwidth]{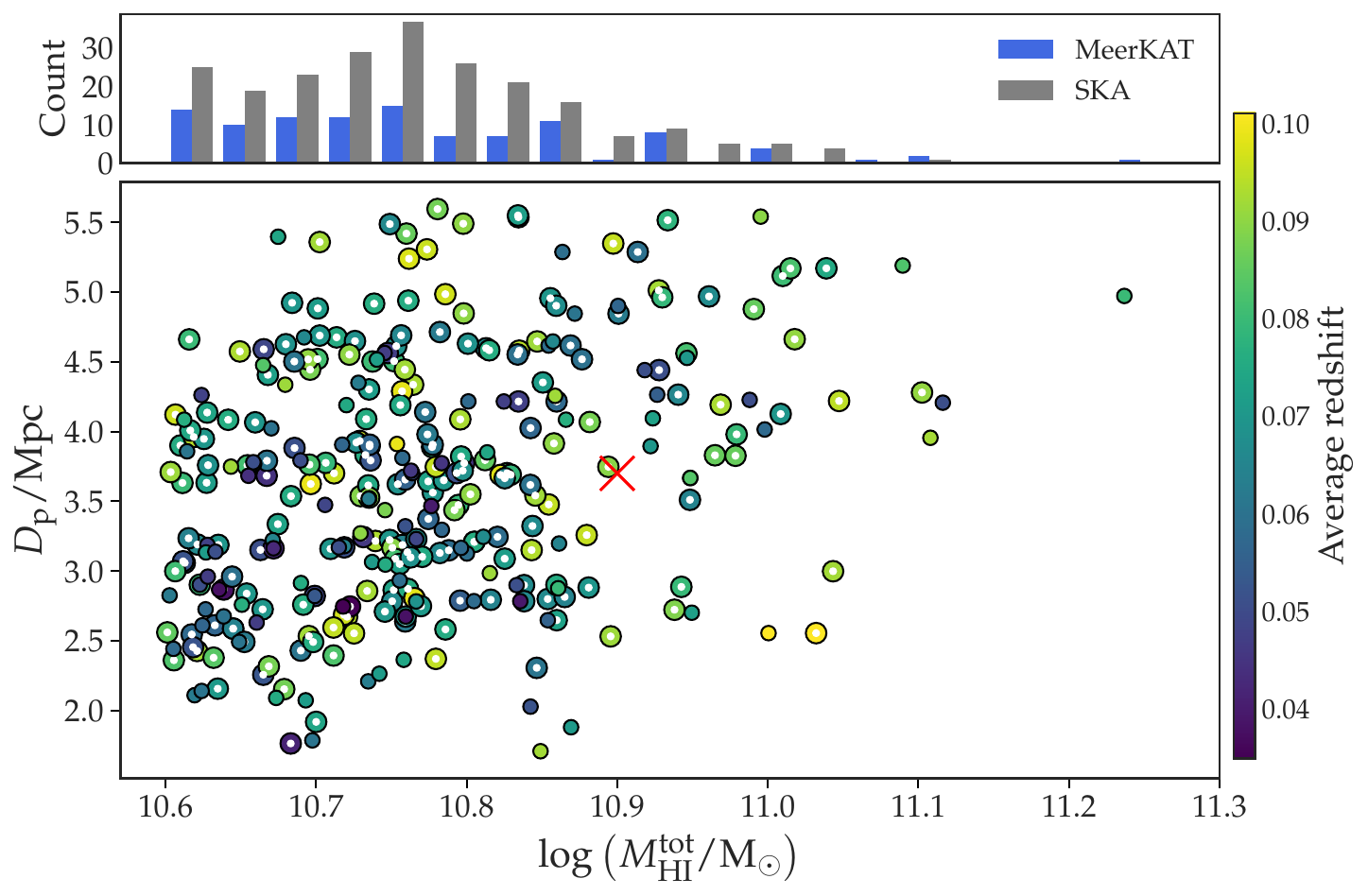}
    \caption{Scatter plot of detectable \textsc{Simba} analogous structures found within 300 light cone realisations where $D_{\rm{p}}$ (projected distance between furthest group members) is plotted as a function of $\log\left(M^{\rm{tot}}_{\rm{HI}}\right)$ (total group \hi mass). Open circles represent SKA-mid-detectable groups, while solid circles represent MeerKAT-detectable groups. All markers are colourised by the mean redshift of the galaxies within the group. The red cross shows the position of the MeerKAT observational structure detected in this work. A histogram is displayed above the plot showing the number of analogues detectable by SKA and MeerKAT for mass bins of $0.035\,\rm{dex}$.}
    \label{fig:proj_vs_mass_sim}
\end{figure*}

Fig~\ref{fig:proj_vs_mass_sim} shows the projected distance between the furthest group members, $D_{\rm{p}}$, as a function of the total \hi mass of the group, $M_{\rm{HI}}^{\rm{tot}}$, for 300 lightcone realisations. Open circles represent SKA-mid-detectable groups, while solid circles represent MeerKAT-detectable groups given the detection criteria outlined above. All markers are colourised by the average redshift of the galaxies within the group. The red cross shows the group studied in this work at $\log (M_{\rm{HI}}^{\rm{tot}}/{\rm{M}_\odot})=10.9$ and $D_{\rm{p}}\sim 3.7\,\rm{Mpc}$. The detection reported in this work lies well within the scatter of the galaxy groups found in \textsc{Simba}, demonstrating that the group studied in this work is not extreme with respect to the simulation.

In 300 realisations of randomly generated lightcones, a mean of 0.37 groupings is associated (and detectable by a MeerKAT observation with similar integration time) per pointing after the selection criteria have been imposed and the FoF algorithm implemented. This together with Fig~\ref{fig:lightcones} and Fig.~\ref{fig:proj_vs_mass_sim} suggests that extended group structures matching our analogue definition are fairly common within the \textsc{Simba} simulation.

\subsubsection{Scaling relations of \textsc{Simba} galaxies}
\label{subsec:simba_sr}

We investigate the position of the simulated \textsc{Simba} galaxies on the \citet{Parkash_2018} scaling relations in Fig.~\ref{fig:hi_vs_sm} and Fig.~\ref{fig:sfr_vs_sm}. The probability density function of the respective data sets is estimated using Gaussian kernel smoothing with Scott bandwidth determination from \textsc{scipy}. Contours are then placed such that the dashed line encloses 68$\,\%$ of the data points based on the distribution estimate. In both figures, the analogue members are galaxies belonging to analogous structures identified within the lightcones as per subsection~\ref{analogue_criteria} and are shown in dashed-line blue contours. The "field galaxies" are randomly selected galaxies within the \textsc{Simba} simulation that meet the detection criteria in subsection~\ref{criteria} and are shown in dashed-line yellow contours.

Generally, the subsets of \textsc{Simba} galaxies probed in this work are in agreement with the $M_{\rm{HI}} \text{--} M_\ast$ relation from \citet{Parkash_2018} as seen in Fig.~\ref{fig:hi_vs_sm}. The most notable feature present is the lobe in the contour of the analogue members in the high-$M_\ast$ regime. This suggests that analogue members are more likely to be found above $\sim \log \left( M_\ast/\rm{M}_\odot \right)=10$ compared to field galaxies. Additionally, the analogue members are on average \hi poor when compared to field galaxies. Similarly, in Fig.\ref{fig:sfr_vs_sm}, the SFMS shows good general agreement between the \citet{Parkash_2018} relation and the \textsc{Simba} galaxies. The same high-$M_\ast$ regime lobe appears in the plot. Again, we see the field galaxies are on average more star-forming than the analogue members indicated by the position of the 68$\,\%$ contours. These trends hint towards evolutionary differences between field galaxies and analogue members, possibly in connection with the enhanced interaction rate observed in denser local environments. In both Fig.\ref{fig:hi_vs_sm} and Fig.\ref{fig:sfr_vs_sm} the most obvious outlier from our observational sample, when compared to the \textsc{Simba} samples, is ID 8, a clearly interacting system of galaxies with an inflated stellar mass.

\section{Conclusions}
\label{sec:conclusion}

We present a multiwavelength analysis of a grouping of 20 \hi-rich galaxies at $z=0.04$ serendipitously discovered with MeerKAT. Ancillary data from DES and WISE are used to enrich the study by enabling identification of optical counterparts using high-resolution \textit{g}-band imaging and derivation of stellar masses and SFRs. This, together with the wealth of \hi information (\hi moment maps, integrated line profiles, and masses), allows for comparison to scaling relations from \citep{Parkash_2018}.

A number of interacting systems and interesting individual galaxies within the system are present in this filamentary-like structure. In particular, we note the presence of a compact group near the projected geometric mean of the group. Within this compact system, we highlight galaxies that appear to be disturbed through tidal interactions with companion galaxies and a potentially ram-pressure-stripped galaxy. Furthermore, many galaxies in the full sample have clear MeerKAT L-band continuum counterparts, which are typically co-located with the galaxy nucleus, indicating potential AGN or star formation activity.

The properties of the structure as a whole are determined and compared to other recent MeerKAT group discoveries such as \citet{Ranchod_2021} and \citet{Glowacki_2024}. Lastly, we place the group into context within the large-scale structure of the universe as well as probe the serendipity of MeerKAT for \hi discovery using the \textsc{Simba} cosmological hydrodynamical simulation, concluding that the discovery is not extreme in the context of the simulation. This suggests the sample size of such objects will grow substantially, allowing for statistical studies of their relation to the large-scale structure, specifically galaxy spin orientation, atomic gas fraction and morphology with respect to dark matter filaments. 

\begin{acknowledgements}
      We thank the anonymous referee for their insightful comments and constructive suggestions, which helped improve the clarity and quality of this work. We would like to thank the staff of the South African Radio Astronomy Observatory (SARAO\footnote{www.sarao.ac.za}) who made these observations possible. We thank Jacques Smulders for his technical assistance with the lightcone production code. We also thank Mpati Ramatsoku and Gyula (Josh) Józsa for helpful discussions. GDL and RPD acknowledge funding from the South African Radio Astronomy Observatory (SARAO), which is a facility of the National Research Foundation (NRF), an agency of the Department of Science and Innovation (DSI). RPD acknowledges funding by the South African Research Chairs Initiative of the DSI/NRF (Grant ID: 77948). The MeerKAT telescope is operated by the South African Radio Astronomy Observatory, which is a facility of the National Research Foundation, an agency of the Department of Science and Innovation. We acknowledge the use of the \textsc{ILIFU}\footnote{www.ilifu.ac.za} cloud computing facility - www.ilifu.ac.za, a partnership between the University of Cape Town, the University of the Western Cape, Stellenbosch University, Sol Plaatje University, the Cape Peninsula University of Technology and the South African Radio Astronomy Observatory. The \textsc{ILIFU} facility is supported by contributions from the Inter-University Institute for Data Intensive Astronomy (IDIA - a partnership between the University of Cape Town, the University of Pretoria and the University of the Western Cape), the Computational Biology division at UCT and the Data Intensive Research Initiative of South Africa (DIRISA). This work has made use of the Cube Analysis and Rendering Tool for Astronomy \citep[CARTA,][]{Comrie_2021}. GDL and RPD acknowledge funding from the Wits Foundation UK in support of travel for research. This research made use of Astropy\footnote{http://www.astropy.org}, a community-developed core Python package for Astronomy. This research makes use of the semi-automated MeerKAT data calibration pipeline, \textsc{Oxkat}\footnote{https://github.com/IanHeywood/oxkat}.
\end{acknowledgements}

\section*{Data Availability}
 
The raw visibilities used in this work can be found at the SARAO archive\footnote{https://archive.sarao.ac.za}. The authors may make data products available upon reasonable request. 

   \bibliographystyle{aa}
   \bibliography{aa55709-25.bib}

\begin{appendix}

\onecolumn
\section{\hi spectra}
\label{app:a}

\begin{center}
    \includegraphics[width=\columnwidth]{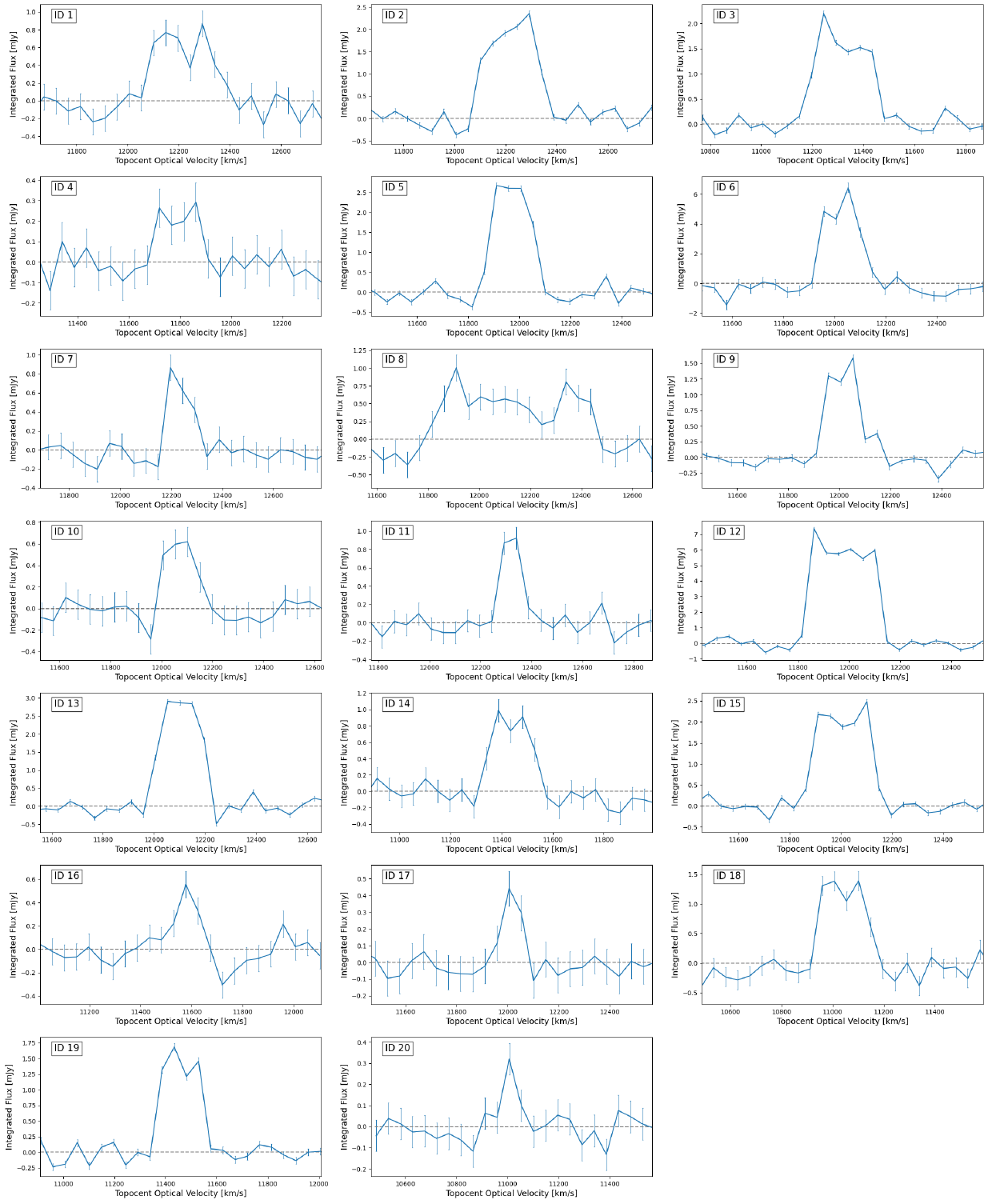}
    \captionof{figure}{Integrated \hi spectra of all group galaxies. Error bars are derived from the per-channel RMS of the local noise within the bounding box of the source.}
    \label{fig:specs}
\end{center}

\begin{landscape}
\section{Multiwavelength imaging of group galaxies/sources}
\label{app:b}
\begin{figure}[h!]
    \includegraphics[height=0.85\textheight]{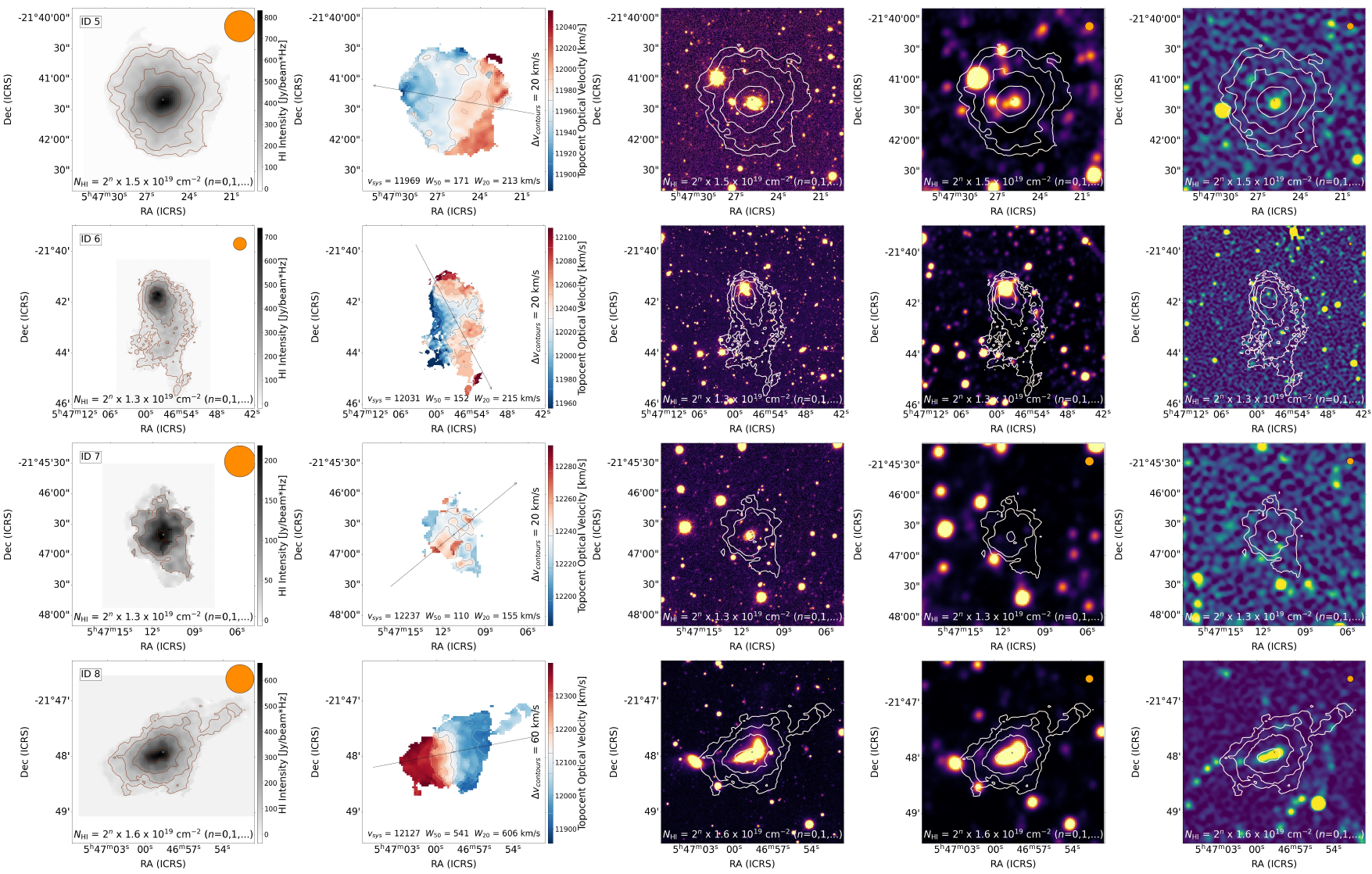}
    \caption{As per Fig.~\ref{fig:page1}, but for source IDs 5-8.}
    \label{fig:page2}
\end{figure}
\end{landscape}

\begin{landscape}
\begin{figure}
    \centering
    \vspace{1cm}
    \includegraphics[height=0.85\textheight]{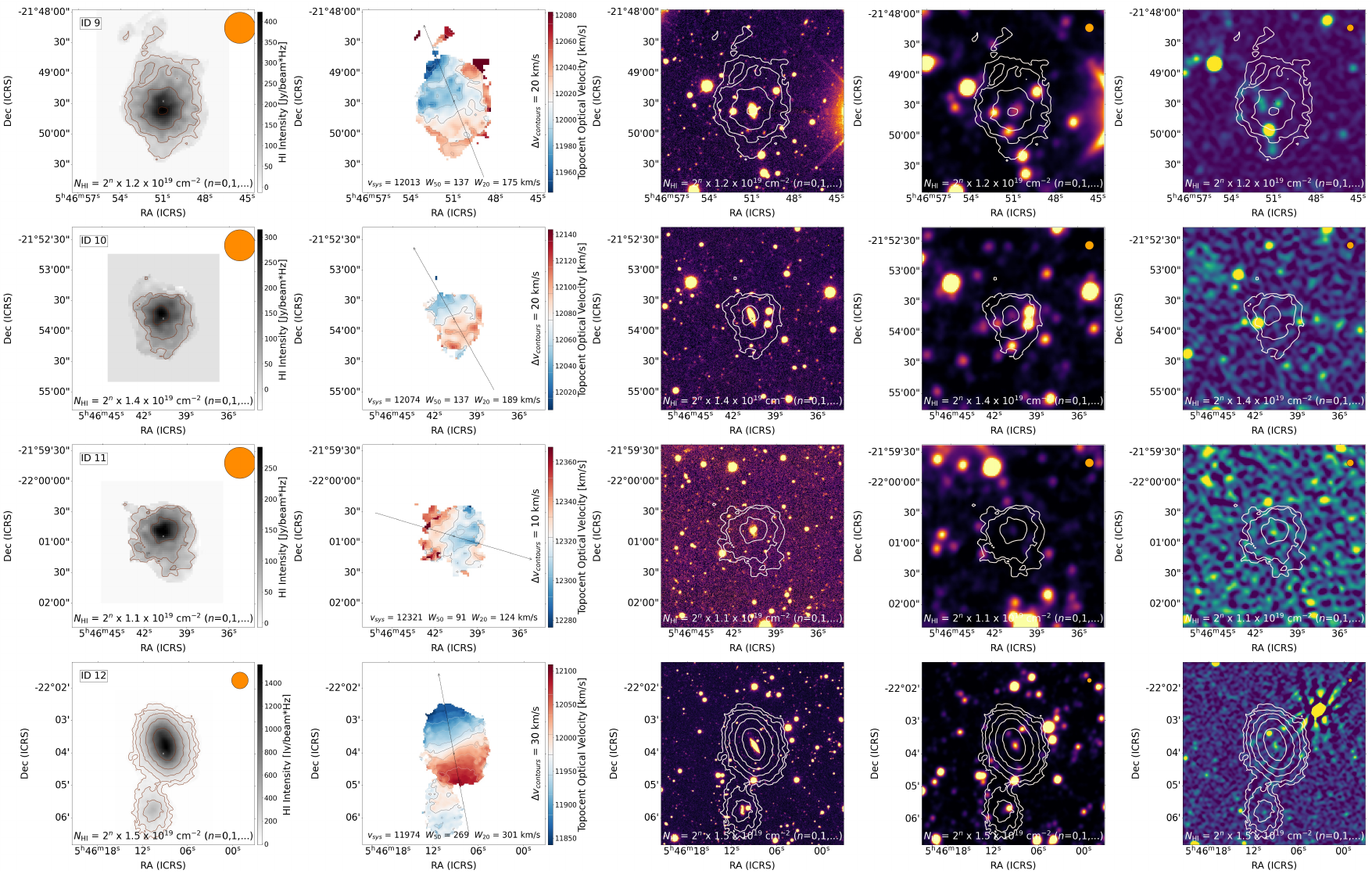}
    \caption{As per Fig.~\ref{fig:page1}, but for source IDs 9-12.}
    \label{fig:page3}
\end{figure}
\end{landscape}

\begin{landscape}
\begin{figure}
    \centering
    \vspace{1cm}
    \includegraphics[height=0.85\textheight]{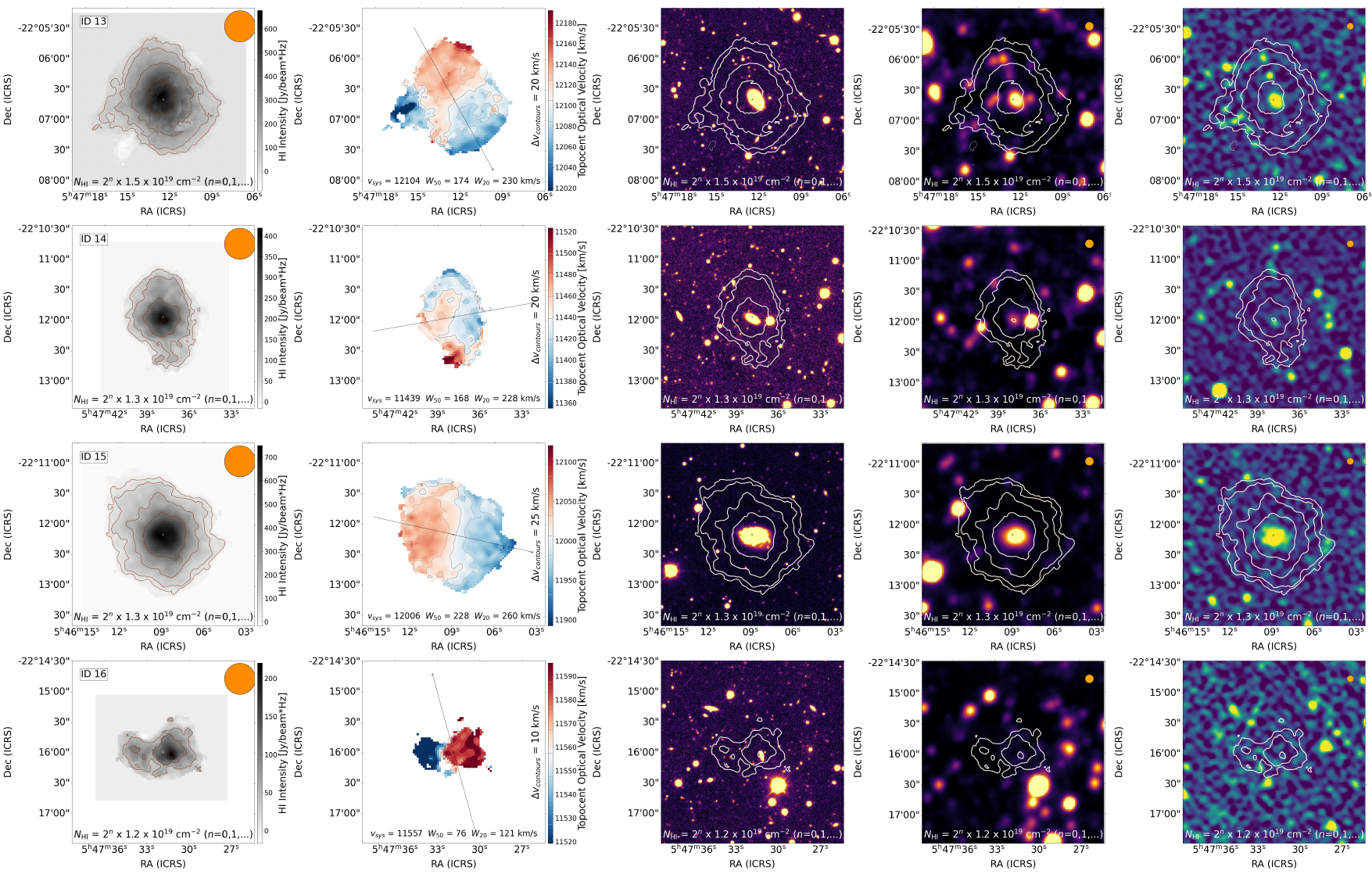}
    \caption{As per Fig.~\ref{fig:page1}, but for source IDs 13-16.}
    \label{fig:page4}
\end{figure}
\end{landscape}

\begin{landscape}
\begin{figure}
    \centering
    \vspace{1cm}
    \includegraphics[height=0.85\textheight]{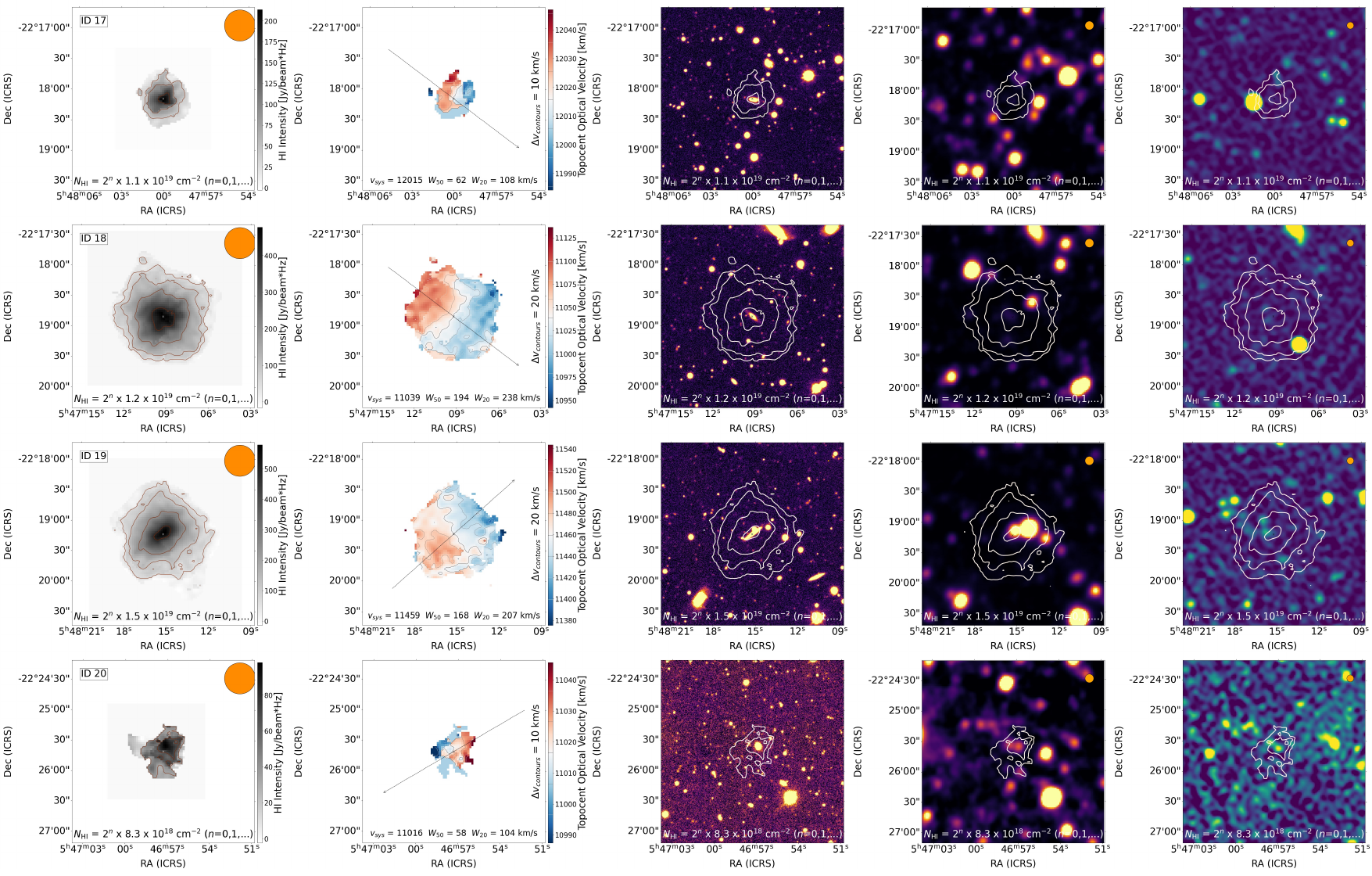}
    \caption{As per Fig.~\ref{fig:page1}, but for source IDs 17-20.}
    \label{fig:page5}
\end{figure}
\end{landscape}

\end{appendix}

\end{document}